\documentclass[showpacs,pra,aps,superscriptaddress,twocolumn,floatfix]{revtex4}
\usepackage{graphicx,float,amsmath,amssymb}
\newfont{\ensmathquatorze}{msbm10 scaled 1400}
\newfont{\ensmathonze}{msbm10 scaled 1100}
\newfont{\ensmathdix}{msbm10}
\newfont{\ensmathneuf}{msbm10 scaled 833}
\newfont{\ensmathhuit}{msbm10 scaled 694}
\newfam\ensmathfam                        
\textfont\ensmathfam=\ensmathonze        
\scriptfont\ensmathfam=\ensmathdix       
\scriptscriptfont\ensmathfam=\ensmathhuit

\renewcommand{\leq}{\leqslant}

\begin{document}

\title{Nonlinear transport of Bose-Einstein condensates through mesoscopic
  waveguides} 

\author{T. Paul}
\affiliation{Laboratoire de Physique Th\'eorique
et Mod\`eles Statistiques, CNRS,
Universit\'e Paris Sud, UMR8626, 91405 Orsay Cedex, France}
\author{M. Hartung}
\affiliation{Institut f\"ur Theoretische Physik,
Universit\"at Regensburg, 93040 Regensburg, Germany}
\author{K. Richter}
\affiliation{Institut f\"ur Theoretische Physik,
Universit\"at Regensburg, 93040 Regensburg, Germany}
\author{P. Schlagheck}
\affiliation{Institut f\"ur Theoretische Physik,
Universit\"at Regensburg, 93040 Regensburg, Germany}

\begin{abstract}
We study the coherent flow of interacting Bose-condensed atoms in mesoscopic
waveguide geometries. 
Analytical and numerical methods, based on the mean-field description of the
condensate, are developed to study both stationary as well as time-dependent
propagation processes. 
We apply these methods to the propagation of a condensate through an atomic
quantum dot in a waveguide, discuss the nonlinear transmission spectrum and
show that resonant transport is generally suppressed due to an
interaction-induced bistability phenomenon. Finally, we establish a link
between the nonlinear features of the transmission spectrum and the
self-consistent quasi-bound states of the quantum dot. 
\end{abstract}


\maketitle

\newcommand{\GP}{Gross-Pi\-ta\-evs\-kii  equation }

\section{Introduction}\label{sec1}
The development of microscopic trapping potentials for ultracold atoms has
lead to a number of fascinating experiments probing the behaviour of
Bose-Einstein condensates on mesoscopic length scales.
Examples include the realization of a Josephson weak link between two
condensates in a double well potential \cite{AnkO05PRL}, the measurement of
interference and phase coherence between two spatially separate condensates
\cite{ShiO05PRA,SchO05NP}, as well as the diffraction of a condensate
from a magnetic lattice \cite{GueO05PRL}.
A convenient setup for such experiments is provided by ``atom chips''
\cite{FolO00PRL} where microscopic confinement potentials are created with the
magnetic field that is induced by current-carrying electric wires mounted on
top of the chip surface.
This technique does not only allow one to produce microtraps, but also to
create
waveguide geometries for cold atoms that can be rather flexible, and thereby
opens the way to explore \emph{transport} properties of cold atomic gases.
Early experiments on atom chips did indeed focus on the propagation of a
Bose-Einstein condensate along such a magnetic waveguide, where the condensate
was transported in a controlled way by means of time-dependent magnetic fields
\cite{HaeO01N} or accelerated along the guide by means of a field gradient
\cite{OttO03PRL}.

The possibility to create such waveguides for cold atoms have stimulated a
number of theoretical investigations on the transport physics of interacting
matter waves, with particular emphasis on possible analogies with mesoscopic
phenomena in the electronic context.
This started with the attempt to define an atomic analog of Landauer's
quantization of the conductance \cite{ThyWesPre99PRL}, and was continued by
the generalization of the ``Coulomb blockade'' phenomenon to cold bosonic atoms
propagating through a quantum-dot-like potential \cite{CarLar99PRL,Car01PRA}.
More recent studies, which are based on an elaborate framework for the
description of scattering processes of Bose-Einstein condensates (to be
described in this article), include the nonlinear resonant transport of a
condensate through atomic quantum dots \cite{PauRicSch05PRL,RapWitKor06PRA},
the manifestation or absence of Anderson localization in the transport through
disorder potentials \cite{PauO05PRA, Pau07PRL}, as well as the transport of
solitons through disorder \cite{BilPav05PRL}.
For their experimental realization, these transport processes would require a
coherent quasi-stationary flow of Bose-Einstein condensed atoms in the
waveguide, which was recently realized in the context of optical guides
\cite{Gue06PRL} using the principle of ``atom lasers'' \cite{BloHaeEss99PRL}.

From the theoretical point of view, the main complication in the description
of a quasi-stationary scattering process of a Bose-Einstein condensate
obviously comes from the presence of the atom-atom interaction.
In leading order, the effect of this interaction is included in a
nonlinear term in the Schr\"odinger-like Gross-Pitaevskii equation for the
condensate wavefunction.
In presence of a waveguide potential, providing a harmonic confinement in two
(transverse) spatial dimensions and permitting free motion along the third
(longitudinal) dimension, an adiabatic treatment of the transverse degrees of
freedom allows one to describe the evolution of the condensate by means of an
effective 
one-dimensional Gross-Pitaevskii equation as long as the confinement
of the waveguide is sufficiently strong (such that the condition for the
``1D mean-field regime'' is satisfied \cite{MenStr02PRA}).
This one-dimensional nonlinear wave equation does permit stationary solutions
corresponding to condensates that propagate with finite velocity along the
axis of the guide \cite{LebPav01PRA}.
As was shown by Leboeuf and Pavloff, these solutions can then be used in order
to construct scattering wavefunctions of the condensate (with the appropriate
outgoing boundary condition) in presence of finite-range perturbation
potentials in the waveguide \cite{LebPavSin03PRA}.

In contrast to the linear Schr\"odinger equation, the knowledge of stationary
scattering states alone does not necessarily permit the prediction of the
outcome of a given propagation experiment with Bose-Einstein condensates.
This is not only the case for the propagation of finite wave packets (which
obviosly cannot be decomposed into individual scattering eigenstates, due to
the absence of the superposition principle in the Gross-Pitaevskii equation),
but applies also to adiabatic injection processes as performed in
Ref.~\cite{Gue06PRL}, where the waveguide is gradually filled with matter
waves.
Clearly, if such an adiabatic process leads to a quasi-stationary flow of the
condensate (which actually need not be the case, as we pointed out in
Ref.~\cite{PauO05PRA}), the corresponding scattering state necessarily
satisfies the stationary Gross-Pitaevskii equation.
However, not every scattering eigenstate of this nonlinear Schr\"odinger
equation  
can eventually be populated in this way:
due to the nonlinearity, the eigenstates in the waveguide can be
\emph{dynamically unstable}, which means that they would disintegrate in the
course of time evolution as a consequence of small deviations.
Such dynamical stability properties cannot easily be inferred from the
stationary 
Gross-Pitaevskii equation.
Another nontrivial problem is, as we shall explain below, the determination of
the incident flux of atoms that is associated with a given stationary
scattering state. 
This information is required in order to establish the connection to a given
propagation experiment (where the incident current is typically under much
better control than the net current during the propagation) and to determine
the transmission coefficient of the scattering state.

In view of these complications, it seems advisable to study waveguide
scattering of Bose-Einstein condensates within the framework of the
\emph{time-dependent} Gross-Pitaevskii equation.
While the straightforward numerical simulation of wave packet propagation
processes is hardly feasible in the limit of spatially broad and energetically
narrow wave packets (which would be required, e.g., for studying the
energy-resolved transmission through atomic quantum dots), it is possible to
directly simulate the quasi-stationary injection process from an external
reservoir of Bose-Einstein condensed atoms into the waveguide, as it was
experimentally performed in Ref.~\cite{Gue06PRL}.
Assuming that this reservoir is sufficiently large such that the effect of the
back-action from the waveguide can be neglected, the dynamics in the waveguide
is effectively described by an \emph{inhomogeneous} Gross-Pitaevskii equation,
which contains a source term that models the input of matter waves from the
reservoir.
This inhomogeneous Schr\"odinger-like equation can be efficiently integrated with
standard finite-difference methods, using absorbing boundary conditions in
order to avoid artificial backreflections from the ends of the numerical grid.
Typically one would start with vanishing condensate density in the guide, and
then time-integrate the equation while adiabatically increasing the source
amplitude from zero up to a given maximal value.
Clearly, this approach is rather close to the realistic experiment.
By construction, it automatically yields, at the end of the propagation,
scattering states that are dynamically stable (provided the flow remains
quasi-stationary during the integration), and it allows in a natural way to
determine the transmission of those states.
We have successfully applied this approach to the transport of Bose-Einstein
condensates through quantum-dot-like double barrier potentials
\cite{PauRicSch05PRL} and through one-dimensional disorder potentials
\cite{PauO05PRA}.

The present paper is devoted to the detailed description of this
time-dependent approach to nonlinear waveguide scattering of a
Bose-Einstein condensate, and to its relation with the existence of stationary
scattering states of the condensate.
To this end we briefly review in Sec.~\ref{sec2} the so called 1D mean-field
regime, set up the theoretical framework to study transport and scattering
processes, and introduce concepts that allow to define transmission and
reflection coefficients for stationary scattering states that are solutions of
a nonlinear wave equation.
In Sec.~\ref{sec2C}, the numerical method that is based on integrating the
time-dependent \GP in presence of a source term is explained.
As a first application, the transmission spectrum of the condensate flow
through a quantum point contact consisting of a single potential barrier in
the waveguide is discussed in Sec.~\ref{sec2D}.
In Sec.~\ref{sec3} we investigate the transport through a symmetric double
barrier potential and we show in \ref{sec3A} that the transmission spectrum
exhibits an interaction induced suppression of resonant transport.
Finally, in Sec.~\ref{sec3C}, we develop an analytical description of the
transport problem through the double barrier potential in terms of internal
quasi-bound states. This establishes a clear link between the nonlinear
signatures of the transmission spectrum and the self-consistent quasi-bound
states of the quantum dot. 

\section{Mean-field approach to transport of condensates}
\label{sec2}

In the following we consider a coherent beam of Bose-Einstein condensed atoms
at zero temperature, propagating through a cylindrical waveguide with a
finite-range scattering potential, given, e.g., by a constriction acting as
a barrier potential for the beam.
One of the aims of this work is to develop new methods to describe such
propagation processes based on the Gross-Pitaevskii mean-field theory
\cite{PitStribook,DalO99RMP}.
The mean-field dynamics of a dilute condensate can be described in terms of a
macroscopic order parameter, the condensate wave function $\Psi(\vec r,t)$, which
obeys the nonlinear \GP \cite{CasDum96PRL} 
\begin{equation}\label{10} 
i\hbar\frac{\partial}{\partial t}\Psi(\vec r,t) = \left[-\frac{\hbar^2}{2m}\Delta +V(\vec r)+
  U_0 |\Psi(\vec r,t)|^2\right]\Psi(\vec r,t).
\end{equation}
Low-energy scattering processes between two atoms in the condensate are
predominately described by the contribution from s-wave scattering and lead to
the nonlinear term $U_0 \vert \Psi(\vec r,t) \vert^2 $. Here $U_0 = 4 \pi \hbar^2
a_s/m$ is the interaction strength which is determined by the s-wave
scattering length $a_s$ and the mass $m$ of the condensed bosons. 
The term $V(\vec r)$ in Eq.~(\ref{10}) is the external trapping potential
experienced by the atoms. 
For the sake of definiteness we consider the experimentally relevant case of a
condensate in a cylindrical harmonic waveguide with an additional scattering
potential that is induced along the guide.
Let $x$ be the coordinate along the axis of the guide and 
$r \equiv \sqrt{x^2 + y^2} $ the cylindrical radius associated with the transverse
coordinates, then we assume $V(\vec r)$ to be of the form 
\begin{equation}\label{20}
V(\vec r)= \frac{1}{2}m \omega^2 r^2 + V_{||}(x).
\end{equation}
Here, the first term on the right-hand side is the transverse harmonic
confinement of the guide with trapping frequency $\omega$ and $V_{||}(x)$
is the scattering potential parallel to the axis of the guide.
$V_{||}(x)$ could, e.g., consist of a single barrier that acts as a
constriction for the condensate flow. Such a barrier can, for instance, be
induced by irradiating a strongly focused blue-detuned laser beam onto
the waveguide.

\subsection{1D mean-field regime}\label{sec2A}

In this subsection we derive an effective one-dimensional version of the \GP
which is particularly suited to describe condensates in elongated waveguide
structures. To this end, we adopt the adiabatic approximation method outlined
in Refs.~\cite{JacKavPet98PRA,MenStr02PRA,LebPav01PRA}, where the condensate
wave function can be cast into the form
\begin{equation}\label{30}
\Psi(x,r) = \psi(x,t)\phi(r, n).
\end{equation}
Here, $\phi$ is the equilibrium ground state wave function for the transverse
motion, normalized to unity 
\begin{equation}\label{40}
\int d^2r|\phi|^2=1,
\end{equation}
$\psi(x,t)$ describes the longitudinal motion, and the density per unit of
longitudinal length is given by 
\begin{equation}\label{50}
n(x,t)\equiv\int d^2r|\Psi|^2=|\psi(x,t)|^2.
\end{equation}
We remark that this adiabatic ansatz involves a local density approximation,
in the sense that one assumes that the transverse motion depends solely on the
local condensate density $n(x,t)$ at position $x$.
It was pointed out in Ref.~\cite{JacKavPet98PRA} that this approximation is
justified if the transverse scale of the density variation is much smaller
than the longitudinal one.
This regime is certainly reached when the scale of variation of the
longitudinal potential $V_{||}(x)$ is considerably larger than the harmonic
oscillator length $a_{\perp}=\sqrt{\hbar/(\omega m)}$ of the radial transverse confinement.

Inserting the ansatz (\ref{30}) into the Gross-Pitaevskii equation (\ref{10})
yields
\begin{eqnarray}\label{55}
i\hbar\phi\frac{\partial}{\partial t}\psi&=&-\phi\frac{\hbar^2}{2m}\psi+\psi\left[ -\frac{\hbar^2}{2 m}\left(
    \frac{\partial^2}{\partial r^2}+ \frac{1}{r}\frac{\partial}{\partial r}  \right) + \right. \nonumber\\ 
 &&\left.  +\frac{1}{2}m\omega^2r^2+U_0 n(x,t) |\phi|^2   \right]\phi.
\end{eqnarray}
We can identify the term in the square brackets as the effective
Hamiltonian $H_{T}$  for the transverse degree of freedom, acting on the wave
function $\phi$, 
\begin{equation}\label{57}
H_{T}\phi = \epsilon(n)\phi.
\end{equation}
The energy $\epsilon(n)$ associated with the transverse state $\phi$ depends
parametrically on the longitudinal density $n$. 
Thus, we obtain a pair of equations, one for the transverse, and one for the
longitudinal dynamics of the condensate, 
\begin{eqnarray}\label{60}
 \epsilon(n)\phi &=&
\left[-\frac{\hbar^2}{2 m}\left( \frac{\partial^2}{\partial r^2}+ \frac{1}{r}\frac{\partial}{\partial r}
  \right)+U_0 n(x,t) |\phi|^2  \right. \nonumber\\  
  & & \left. + \frac{1}{2}m\omega^2r^2 \right]\phi   ,\label{70}   
 \\
i \hbar \frac{\partial}{\partial t}\psi &=& \left[-\frac{\hbar^2}{2 m} \frac{\partial^2}{\partial
    x^2}+V_{\|}(x)+\epsilon(n(x,t))\right]\psi \label{80}. 
\end{eqnarray}
Eq.~(\ref{80}) is an effective one dimensional wave equation for the
longitudinal order parameter $\psi$ which is particularly suited to describe a
condensate in non-uniform waveguides. This regime is often denoted as the 1D
mean-field regime \cite{MenStr02PRA}. 

It remains to determine $\epsilon(n)$.
In the following, we assume that $\phi$ is the energetic ground state of $H_T$.
In the so-called low density limit, $a_s n \ll 1$, the nonlinear term $U_0 n
|\phi|^2$ in Eq.~(\ref{70}) is a small perturbation and a first-order
perturbative  solution  of Eq.~(\ref{70}) yields
\begin{eqnarray}\label{90}
\epsilon(n)=\epsilon_0 + U_0 \langle \phi \left| |\phi_0 |^2 \right| \phi_0 \rangle = \epsilon_0  +  2\hbar \omega a_s n,
\end{eqnarray}
where $\epsilon_0 = \hbar \omega$ is the eigenenergy of the ground state $\phi_0$ of the
unperturbed transverse Hamiltonian ($\epsilon_0$ is a constant energy shift which we
drop in the following).
In the opposite large density limit, $a_s n \gg 1$, the kinetic energy term in
Eq.~(\ref{80}) can be neglected, and the so called Thomas-Fermi approximation
holds for the transverse wave function \cite{DalO99RMP}, yielding
\begin{equation}\label{100}
\phi_{TF}= \frac{1}{U_0\sqrt{n}}\sqrt{\epsilon(n)-V_{\perp}(r)} \ .
\end{equation}
By imposing the normalization condition (\ref{40}) to the Thomas-Fermi wave
function (\ref{100}) we find in the high-density regime
\begin{equation}\label{110}
\epsilon (n)=2 \hbar \omega \sqrt{n a_s}\ .  
\end{equation}

At this point we remark that the validity of the \GP is restricted to the
dilute gas regime, where the 3D density $n_{3d}$ fulfills $n_{3d}a^3_s \ll 1$
\cite{PitStribook,DalO99RMP}.  
This condition reads in the 1D mean-field regime $n a_s \ll (a_{\perp}/a_s)^{2 / \nu
}$ ($\nu = 1$ in the low density regime and $\nu = 1/2$ for high densities
\cite{LebPav01PRA}). 
Typically $a_{\perp}/ a_s$ is of the order $10^3$. This condition will be
considered as always fulfilled, even in the regime of high longitudinal
densities, when $n a_s \gg 1$. 
On the other hand, the weakly interacting 1D Bose gas picture also breaks down
at very low densities, in the Tonks-Girardeau regime (see e.g. 
Refs.~\cite{PetShlWal00PRL,DunLorOls01PRL,Ols98PRL,Thy99}). This occurs in the
regime $n a_s \ll (a_s / a_{\perp})^2 \simeq 10^{-6} $ which we therefore discard from
our present study. 

\begin{figure}[ptb]
\centering
\includegraphics[width=0.9\linewidth,angle=0]{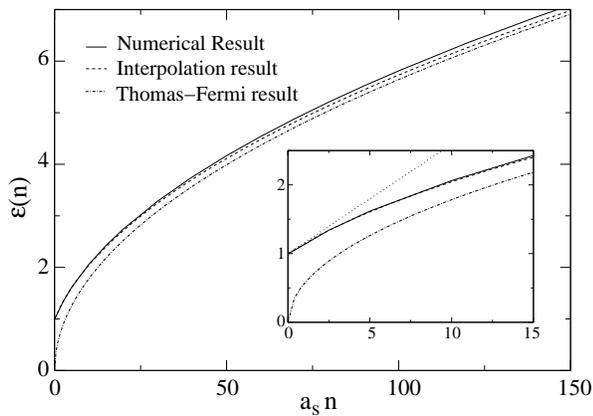}
\\[4mm]
 \caption{\label{Fig_epsilon} Transverse energy $\epsilon(n)$ (in units of $ \hbar \omega$)
   as a function of $a_s n$ (in units of $\hbar^2/m$). The numerical result (solid
   line) coincides for large values $a_s n$ very well with  the Thomas-Fermi
   result. The interpolation result agrees excellently with the numerical
   result for small values of $a_s n$ and converges towards the Thomas-Fermi
   result for large $a_s n$.
The inset zooms into the region of small $a_s n$. The straight dashed line
displays the perturbative result (\ref{90}). 
}
\end{figure}

At the end of this section, we derive an analytical expression that allows to
interpolate $\epsilon (n)$ between the two opposite limits $n a_s\ll 1$ and $n a_s\gg 1$. 
To this end we consider the ansatz
\begin{equation}\label{120}
\epsilon(n)=\left[\alpha +\beta  (a_s n) + \gamma  (a_s n)^2 \right]^{1/4}.
\end{equation}
To determine the coefficients $\alpha, \beta$ and $\gamma$, we expand Eq.~(\ref{120}) in the
limit $a_s n \ll 1$ to first order in $a_s n$, 
\begin{equation}\label{130}
\epsilon(n)= \alpha^{1/4} + \frac{1}{4}\alpha^{-3/4} \beta (a_s n), \qquad  \ (a_s n)\ll 1.
\end{equation}
In the limit $a_s n \gg 1$ we keep only the  quadratic term $(a_s n)^2$ in
Eq.~(\ref{120}), 
\begin{equation}\label{140}
\epsilon(n)= \gamma^{1/4} \sqrt{a_s n}, \qquad  \ (a_s n)\gg  1.
\end{equation}
The comparison of Eqs.~(\ref{130},\ref{140}) with Eqs.~(\ref{90},\ref{110})
yields $\alpha=\hbar^4\omega^4$, $\beta = 8 \hbar^4\omega^4$ and $\gamma =16 \hbar^4\omega^4$, and the interpolation
formula (\ref{120}) reads 
\begin{equation}\label{150}
\epsilon(n)= \sqrt{\hbar^2 \omega^2 + 4 \hbar^2 \omega^2 (a_s n)} .
\end{equation}
This result can be compared with numerically computed values for $\epsilon(n)$
\cite{Note1}. Indeed, as displayed in Fig.~\ref{Fig_epsilon}, we find a
good agreement between the interpolation result and the numerically computed
values for the whole range of values of $a_s n$ in between the two opposite
density limits.

\subsection{Scattering states in waveguides}\label{sec2B}
In this subsection we study the stationary transport modes of a coherent
condensate flow through a quasi one-dimensional waveguide with a scattering
potential in the 1D mean-field regime. 
Starting point of our considerations is the effectively one-dimensional \GP
(\ref{80}).  
To determine its steady solutions, we write $\psi(x,t)=A(x)\exp[iS(x)]\exp(-i\mu
t)$, where $A(x)$ and $S(x)$ are real valued functions. The longitudinal
density is $n=A^2$, $\mu$ is the chemical potential of the condensate and $v =
(\hbar/m)(dS/dx)$ its local velocity. From Eq.~(\ref{80}) we obtain flux
conservation $n(x) v(x)\equiv j_t = \text{const}$, and an equation of motion for
the amplitude $A(x)$ of the wave function 
\begin{eqnarray}\label{160}
\mu A = -\frac{\hbar^2}{2 m}A'' + \frac{m}{2}~\frac{j_t^2}{n^2} A +V_{{\|}}(x)A+\epsilon(n) A.
\end{eqnarray}
In the following, we assume that the longitudinal potential $V_{||}(x)$
vanishes asymptotically in the ``upstream'' region, i.e.\ for $x \to - \infty$,
and in the ``downstream'' region, for $x \to + \infty$: $V_{||}(x \to \pm \infty) = 0$. 
In accordance with this terminology, we consider an incident beam of
condensate that propagates from $x \to - \infty$ to $x \to + \infty$ (i.e., from the
upstream to the downstream region).

In order to properly define the scattering problem, we first study the
asymptotic behavior  
of the flow far away from the constriction, where $V_{||}(x)=0$.
In this region, Eq.~(\ref{160}) can be integrated once, yielding the first
order equation of motion 
\begin{eqnarray}\label{170}
E &= &\frac{\hbar^2}{2 m} (A')^2 +\frac{m~ j_t^2}{2 A^2} + \mu A^2 - {\mathcal{E}}(n),\\
&& \text{with} \ \ {\mathcal{E}}(n) = \int_0^n \epsilon(\tilde n)d \tilde n,
\end{eqnarray}
where $E$ is an integration constant. 
It was pointed out in Ref.~\cite{LebPav01PRA} that Eq.~(\ref{170}) admits a
simple interpretation in terms of classical dynamics, since it describes the
energy conservation of a fictitious classical particle with ``position'' $A$
and ``time'' $x$ moving in the effective potential  
\begin{eqnarray}\label{173}
W(n)\equiv ({m~ j_t^2})/({2 n}) + \mu n - {\mathcal{E}}(n) \, ,
\end{eqnarray}
and the integration constant $E$ corresponds to the total energy of the
particle. 
Eq.~(\ref{170}) is therefore integrable by quadrature (see Ref.
\cite{RapWitKor06PRA} for a discussion of the low density regime $a_s n \ll 1$). 

\begin{figure}[ptb] 
\centering
\includegraphics[width=1\linewidth,angle=0]{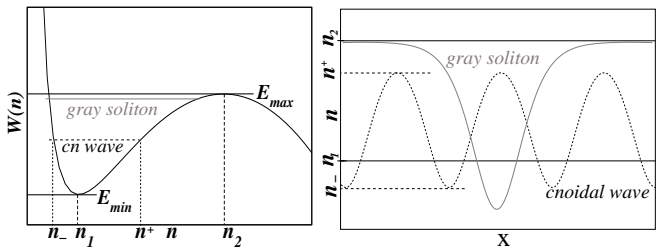}
\\[4mm]
 \caption{\label{Fig_modes} Plot of the function $W(n)$, here displayed for
   the low density regime ${\mathcal E}(n)=g\,n^2/2$, in the left panel. For
   given $\mu, j_t$ and $g$, a beam of uniform density has either a density
   $n_1$ (supersonic solution) or a density $n_2$ (subsonic solution). At a
   given classical energy $E$ (with $E_{min}<E<E_{max}$, to assure bounded
   density oscillations ), $n_-$ and $n^+$ are the minimum and maximum values
   of the cnoidal density oscillations, displayed in the right panel. Energy
   values $E$ close to (but lower than) $E_{max}$ correspond to gray solitons.
}
\end{figure}

The left panel of Fig.~\ref{Fig_modes} displays the potential $W(n)$ in the
low-density regime where we have ${\mathcal E}(n)=g n^2/2$ with the effective
interaction parameter $g\equiv 2\hbar \omega a_s n^2$. $W(n)$ has qualitatively the same
form in the high-density regime as well where ${\mathcal E}$ is given by
${\mathcal E}(n)=2 {\bf g} n^{3/2}/3$ with ${\bf g} \equiv 2\hbar \omega \sqrt{a_s}$ .  
For weak and moderate coupling constants $g$ (respectively ${\bf g}$), $W(n)$
exhibits a local minimum $E_{min}=W(n_1)$ 
at a low density $n_1$ and a local maximum $E_{max}=W(n_2)$ at a high density
$n_2$.
These extrema, at which the fictitious particle would be at rest forever,
correspond to solutions of Eq.~(\ref{170}) with constant density.
They represent plane waves of the form 
$\psi_\nu(x,t)=\sqrt{n_{\nu}}\exp(ik_{\nu}x-i\mu t/\hbar)$ ($\nu = 1,2$) whose wave numbers
$k_{\nu}$ are implicitly determined through the dispersion relation of the \GP,
namely
\begin{eqnarray}\label{180}
\mu=\frac{m}{2}\frac{j_t^2}{n_{\nu}^2}+\epsilon( n_{\nu}),
\end{eqnarray}  
as expressed in terms of the total current $j_t =\hbar k_{\nu} n_{\nu}/m$.
The solutions $\psi_1$ and $\psi_2$ are termed ``supersonic'' and ``subsonic'',
respectively, since the beam velocity is larger than the speed of sound of the
condensate for $\psi_1$ and smaller than the speed of sound for $\psi_2$
\cite{LebPav01PRA}.
The transport of particles at theses two solutions is dominated by the kinetic
energy in the supersonic case, and by the interaction between the atoms in the
subsonic case.
We note that in the noninteracting limit, where $\epsilon(n)$ is independent of $n$,
the subsonic density $n_2$ diverges and $W(n)$ has only one finite extremum at
the density $n_1$.

Solutions of Eq.~(\ref{160}) with $E_{min} < E < E_{max}$ exhibit periodic
density oscillations and correspond to a bounded motion of the fictitious
classical particle. 
They are implicitly given through the integration of
Eq.~(\ref{170}), i.e. 
\begin{eqnarray}\label{185}
x-x_0=\int_{A(x_0)}^{A(x)} \frac{\sqrt{2/m} \ \hbar \ d A}{E-m \, j_t^2
  / (2A^2) - \mu A^2+{\mathcal{E}}( A^2 ) }, 
\end{eqnarray} 
where the amplitude $A(x_0)$ at the position $x_0$ determines the initial
value for the solution of the differential equation (\ref{160}). For $\epsilon(n)=g\,
n/2$ it was shown that the solutions of Eq.~(\ref{185}) can be expressed in
terms of Jacobi-Elliptic functions \cite{RapWitKor06PRA}.
 
For our purpose, a qualitative characterization of the free solutions of the
Gross-Pitaevskii equation is sufficient: 
Small deviations from the constant density value $n_1$, e.g. small values of
$E-W(n_1)$ correspond to small sinusoidal density oscillations.  
Energy values close to (but lower than) the limiting classical energy value
$E_{{max}}=W(n_2)$ correspond to  gray solitons.  
In the intermediate regime, between the limiting cases of small sinusoidal
oscillations and gray solitons, the condensate density exhibits cnoidal
oscillations. 
Energy values larger than  $E_{\text{max}}$ lead to an infinite density at
finite $x$ and cannot be interpreted as physically meaningful steady-state
solutions. 
We also note that the flat-density solutions coincide, $n_1=n_2$, when the
potential $W(n)$ exhibits a saddle point configuration.
For the potential $W(n)$ displayed in Fig.~\ref{Fig_modes}, such a saddle
point configuration would, e.g., be encountered by increasing $g$ while $\mu$
and $j_t$ are kept fixed. 
In the low density limit where $\epsilon(n) = 2\hbar\omega_{\perp}a_s n$, the criterion for the
existence of a saddle point configuration reads $8\mu^3=27m j^2 g^2$; in the
high density limit where $\epsilon(n) = 2\hbar\omega_{\perp} \sqrt{a_s n}$, we find $\mu^5=5^5 m j^2
{{\bf g}}^4  / 2^9$.  
Beyond these limits no stationary solutions exists any more. 

Finding stationary scattering states in presence of a finite scattering
potential requires now to match two asymptotic density modes, each
characterized by a separate integration constant $E$, in the upstream
respectively downstream region.
From general arguments on the dispersion relation of elementary excitations of
the Gross-Pitaevskii equation follows that the physically meaningful boundary
condition for the steady-state solutions of Eq.~(\ref{160}) demands a constant
downstream density profile \cite{LebPav01PRA}. The asymptotic downstream
density should therefore correspond either to $n_1$ or $n_2$. 
In the present study, we intend to investigate the crossover from a
noninteracting to a weakly or moderately interacting system. We therefore
focus on the regime of rather small condensate densities, respectively weak
atom-atom interactions ($n a_s \ll 1$ and $\epsilon (n)=g n$); hence, the low-density
downstream solution $n_1$ will be relevant in the following. 
The high-density solution $n_2$ exhibits qualitatively different features,
such as solitonic transmission modes, and has been discussed in
Ref.~\cite{LebPavSin03PRA}.

In analogy with the scattering problem in a non-interacting system we define a
stationary scattering state as a solution of Eq.~(\ref{80}) of the form  
\begin{eqnarray}\label{186}
\psi(x,t)=\psi(x)\exp(-i\mu t /\hbar),
\end{eqnarray}
satisfying, in the downstream region, outgoing boundary conditions of the
form $\psi(x) =  \sqrt{n_1} \exp(ik x)$, with $k>0$ given by $k_1$ as defined
above. 
In order to determine the scattering states for a given barrier potential
$V_{||}(x)$, which vanishes at $x\to \pm \infty$, and for given values for the total
current flow $j_t$ and the chemical potential $\mu$, we integrate the equation
of motion (\ref{160}) from the downstream to the upstream region with the
``asymptotic condition'' $A=\sqrt{n_1}$ and $A'=0$ in the downstream region.
This allows us to compute the density profile in the whole waveguide, and by
computing the phase via $S'(x) = m j_t A^2(x) / \hbar$ we  determine unambiguously
the stationary scattering state $\psi(x)$. This procedure describes the
scattering process in terms of a so-called {\em fixed output problem}, because
the outgoing current $j_t$ in the downstream region enters as a parameter in
the asymptotic boundary conditions that determine the scattering state
\cite{GreKiv92PR,KnaPapWhi91JSP}. 

There is only a small number of potential configurations, such as the square
well or delta-peak barriers, for which the integration can be carried out
analytically \cite{RapWitKor06PRA}. For the general case, it is convenient to
rewrite Eq.~(\ref{160}) in terms of Hamilton-like equations of motion  
\begin{eqnarray}\label{190}
A'= \frac{\partial {\mathcal H}}{\partial p}&=&\frac{m}{\hbar^2}p,
\nonumber \\
p' = -\frac{\partial {\mathcal H}}{\partial A}&=&
\left( \frac{m j_t^2}{A^4} - 2[\mu - V_{\|}(x) - \epsilon(n)] \right) A,
\end{eqnarray}
where we introduced the canonical momentum $p \equiv (\hbar^2/m)A'$.
These equations of motions can be deduced from the classical Hamiltonian
\begin{eqnarray}\label{200}
{\mathcal H}(A,p)=\frac{\hbar^2}{2 m}p^2+\frac{m j_t^2}{2
  A^2}+[\mu-V_{\|}(x)]A^2-{\mathcal{E}}(A^2). 
\end{eqnarray}
In the picture of the fictitious classical particle, $V_{||}(x)$ plays the
role of a driving force which drives the particle away from the
minimum of the classical potential $W(n)$. 
The classical energy, which is $E_d=W(n_1)$ in the downstream region, is
altered by the amount  
\begin{eqnarray}\label{210}
\Delta E = \int _{-\infty}^{+\infty} V_{||}(x)A(x)A'(x)dx,
\end{eqnarray}
which yields the new classical energy value $ E_u=E_d+\Delta E$ for the
upstream region. 
The asymptotic behavior of the scattering state is then fully determined by
$E_u$ and $E_d$.
The energy transfer $\Delta E$ is a measure for the amplitude of the density
oscillations in the upstream region, i.e. increasing values of $\Delta E$ imply a
larger backreflection. 

Our purpose is now to determine the reflection and transmission coefficients,
$T$ and $R$, associated with the stationary scattering states. 
These quantities are naturally given by $T=j_t/j_i$ and R=$j_r/j_i$, where
$j_i$, $j_t$, and $j_r$ respectively denote the incident, transmitted, and
reflected current of the condensate. The determination of $j_i$
and $j_r$, however, is a nontrivial task, since we can not simply decompose
the upstream wave function into an incident and reflected plane wave component
due to the nonlinearity of the Gross-Pitaevskii equation which does not permit
the application of the superposition principle. We show now how the incident
and reflected currents can nevertheless be defined and calculated in a
meaningful way.

First, we briefly recall a method that has been suggested in
Ref.~\cite{LebPavSin03PRA} and was successfully applied in
Ref.~\cite{PauO05PRA}.
It allows one to determine approximate values for $T$ and $R$ in the regime of
small backreflections or small nonlinearities, by means of an approximate
decomposition of the upstream density into an incident and reflected beam.
We consider here the low density regime $a_s n \ll 1$, e.g. $\epsilon(n) = g n$.
In the upstream region, $n(x)=A^2(x)$ obeys the equation [see Eq.~(\ref{170})]
\begin{eqnarray}\label{220}
  E_u = \frac{\hbar^2}{2 m} \left( \frac{d\sqrt{n}}{dx}\right)^2 + W(n)
\end{eqnarray}
with
\begin{eqnarray}\label{221}
  W(n)=\frac{m~ j^2}{2 n} + \mu n - \frac{1}{2}\,g n^2.
\end{eqnarray}
We write the density in the form $n(x) = n_1 + \delta n(x)$, where $\delta n(x)$
represents the density oscillations originating from back-reflections.
Inserting this ansatz into Eq.~(\ref{220}) and introducing a new effective
wave number 
\begin{eqnarray}\label{225}
  \kappa=k\sqrt{1-{1}/({2\, \xi^2 k^2})}
\end{eqnarray}
(here, $\xi=\hbar/\sqrt{2mn_1 g}$ is the condensate's healing length in the
downstream region) and the characteristic scale 
$\delta n_1= {m}[E_u-W(n_1)]/({\hbar^2 \kappa^2})$ for density oscillations, we obtain 
\begin{eqnarray}\label{230}
\left( \frac{d \delta n}{d x}  \right)^2 + 4\kappa^2\delta n^2=8\kappa^2\delta n_1(n_1+\delta n)+ \frac{4 m
  g }{\hbar^2} \delta n^3
\end{eqnarray}
as an equation of motion for $\delta n(x)$.

Until now, no approximation has been made. In the regime of small
back-reflections, where $| \delta n| / n_1 \ll 1$ holds, or small interaction
parameters $g$ (both limits are covered by the condition $| \delta n| / n_1\ll \kappa^2
\xi^2$, see Ref.~\cite{LebPavSin03PRA}), we neglect the cubic term in
Eq.~(\ref{230}). Thus,
the equation of motion (\ref{230}) corresponds to the dynamics of a shifted
harmonic oscillator, and its solution is given by 
\begin{eqnarray}\label{240}
n(x) = n_1 + \delta n_1 + \sqrt{2 n_1 \delta n_1 +(\delta n_1)^2} \cos(2\kappa x+\theta),
\end{eqnarray}
where $\theta $ is an arbitrary  phaser. The density profile (\ref{240}) is
equivalent to that of the two counterpropagating plane waves with wave vektor
$\kappa$
\begin{eqnarray}\label{250}
\psi_i(x)&=&\sqrt{n_1+\frac{\delta n_1}{2}}~\exp(i\kappa x),
\nonumber \\
\psi_r(x)&=& \sqrt{\frac{\delta n_1}{2}}~\exp(-i\kappa x+i\theta),
\end{eqnarray}
yielding
\begin{eqnarray}\label{251}
n(x) = |\psi_i(x)+\psi_r(x)|^2.
\end{eqnarray}
Identifying  $\psi_i(x)$ as an incident and $\psi_r(x)$ as a reflected wave
component allows one to determine the transmission and reflection coefficients
through 
\begin{eqnarray}\label{260}
T = \left(1+\frac{\delta n_1}{2\,n_1}   \right)^{-1},\quad
R =\left( 1+ \frac{2 n_1}{\delta n_1} \right)^{-1}.
\end{eqnarray}

The approximate nature of Eq.~(\ref{260}) becomes evident if we consider the
conservation of currents.
Computing the incident and reflected current components in the upstream region
yields 
\begin{eqnarray}\label{270}
j_i= \left(n_1+\frac{\delta n_1}{2}\right)\,\frac{\hbar \, \kappa}{m}\, ,\quad
j_r= \frac{\delta n_1}{2}\,\frac{\hbar \, \kappa}{m}\, ,
\end{eqnarray}
whereas we find from the asymptotic downstream behavior of the wave function
the transmitted current component $j_t = n_1 \hbar k /m$. It is easy to see that
the relation $j_t+j_r=j_i$ is exactly fulfilled only in the case of vanishing
atom-atom interactions, i.e.\ $k=\kappa$. In the regime of weak interactions
deviations from the current conservation are of the order 
${\mathcal O}[(k\xi)^{-2}]$
and the approximate approach becomes inappropriate for strong interactions or
large backreflections. 

\begin{figure}[ptb]
\centering
\includegraphics[width=0.9\linewidth,angle=0]{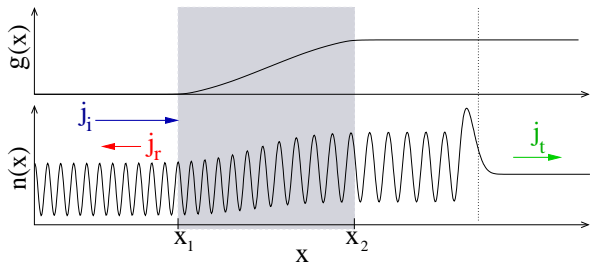}
\\[4mm]
 \caption{\label{Fig_Adpass} (color online) Adiabatic transition of the
   interaction parameter $g$ for a proper definition of transmission
   coefficients: The upper part of the figure displays the adiabatic variation
   of the position-dependent parameter $g(x)$ from $g=0$ up to a maximal value
   $g$. The gray-shaded transition region between $x_1$ and $x_2$ in which
   $g$ varies with position is assumed to be much larger than the typical
   periodicity of the condensate density oscillations. The lower part shows
   the density of a stationary scattering state in presence of a potential
   barrier, which is constant in the downstream region and displays
   oscillations in the upstream region (the position of the barrier potential 
   is marked by the vertical line). The nonlinear cnoidal oscillation of $n$
   between the barrier and $x_2$ is adiabatically conveyed, in the transition
   region between $x_1$ and $x_2$, into a sinusoidal oscillation in the
   interaction-free domain on the left-hand side of $x_1$. There, the wave
   function can be linearly decomposed into an incident and a reflected
   component.
}
\end{figure}

In order to overcome this problem, we consider a waveguide configuration in
which the interaction strength $g$ tends to zero for $x\to -\infty$ and reaches a
finite constant value in the region where the barrier potential is located
(see Fig.~\ref{Fig_Adpass}).
We furthermore assume that the typical length scale on which $g$ varies is
much larger than the periodicity of the density oscillations. 
Such a variation of $g$ can e.g.\ be achieved by decreasing the transverse
confinement frequency $\omega$ of the waveguide or by tuning the scattering length
$a_s$ via a Feshbach resonance.

Using once more the analogy with the dynamics of a classical particle, we
introduce the effective ``pseudo action''
\begin{eqnarray}\label{280}
{\mathcal J} = \oint p\, dA = \frac{\hbar^2}{m}\int _{x_0}^{x_0+\Delta x} [A'(x)]^2 dx
\end{eqnarray}  
that is integrated over one spatial period $\Delta x$ of the upstream density
oscillation (which would be given by $\Delta x = \pi / k$ in the absence of the
interaction).
By use of Eq.~(\ref{170}) the pseudo action can also be written in the form
\begin{eqnarray}\label{290}
{\mathcal J} =
\hbar \, \sqrt{  \frac{2}{m}
}\,\int_{{n_-}}^{{n_+}}\sqrt{[E_u-W(n)]/{n}\,}\, dn,
\end{eqnarray}  
where ${n_-}$, (${n_{+}}$) is the minimal (maximal) density value of the
oscillating upstream density. It is determined via the relation
$W(n_{\pm})=E_u$.
Due to the theorem of adiabatic invariants, ${\mathcal J}$ remains
approximately constant along the waveguide as long as $g$ is sufficiently
slowly varied. It can, under this condition, therefore be evaluated at any
position $x$, in particular also in the far-upstream region at $x < x_1$ where
we have $g = 0$.
There we can decompose the wave function in an incident and reflected part as
\begin{eqnarray}\label{292}
\psi(x) = ( \alpha \text{e}^{i (k x + \varphi)} + \beta \text{e}^{-i k x} ) \text{e}^{i \phi} 
\end{eqnarray}  
with $k=\sqrt{2m \mu}/\hbar$, where the amplitudes $\alpha, \beta$ and the phases $\varphi,\phi$ are
real. The wave function's amplitude reads 
\begin{eqnarray}\label{300}
A(x)=\sqrt{\alpha^2+\beta^2+2\alpha\beta \cos(2kx + \varphi)}, 
\end{eqnarray}
and the canonical momentum $p$ is given by
\begin{eqnarray}\label{310}
p(x)=\frac{\hbar^2}{m}A'(x)=\frac{2\alpha\beta k\sin(2kx + \varphi)}
{\sqrt{\alpha^2+\beta^2+2\alpha\beta \cos(2kx + \varphi)}}.
\end{eqnarray}

By use of $dA=A'dx$, we evaluate Eq.~(\ref{280}) as
\begin{eqnarray}\label{320}
{\mathcal{J}}= \oint p\, dA, = 2\beta^2\,\hbar^2 k \pi / m .
\end{eqnarray}
Using the fact that the incident and reflected currents read $j_i=\hbar k \alpha^2/m$ 
and $j_r=\hbar k \beta^2/m$ in the far-upstream region,
we can obtain the reflection and transmission coefficients via
\begin{eqnarray}\label{330}
R &=& \frac{j_r}{j_i} = \frac{\beta^2}{\alpha^2}=\left( 1+\frac{2 \pi \hbar
    j_t}{{\mathcal{J}}} \right),\nonumber\\ 
T&=& 1 - R = 1-\frac{\beta^2}{\alpha^2}  =\left(1+\frac{{\mathcal{J}}}{2 \pi \hbar
    j_t}\right)^{-1}. 
\end{eqnarray}
Eq.~(\ref{330}) unambiguously assigns a reflection and a transmission value to
each scattering state that is a solution of the nonlinear wave equation
(\ref{80}). This definition represents a natural extension of the concept of
transmission for nonlinear scattering problems.

\begin{figure}[ptb]
\centering
\includegraphics[width=1.0\linewidth,angle=0]{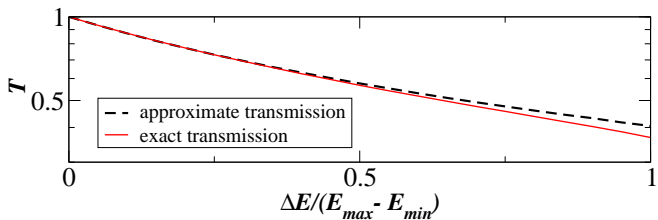}
\\[4mm]
 \caption{\label{Fig_Transvergleich} (color online) Comparison between the
   approximate and exact transmission values, calculated with Eqs.~(\ref{330})
   and (\ref{260}) respectively, for a moderately interacting condensate that
   propagates through a potential barrier, with the dimensionless parameters
   $\mu=3$, $g=1/2$, $j_t=1$, and with variable barrier height. The transmission
   is plotted as a function of the classical energy transfer $\Delta E$, which is a
   measure for the back-reflection in the upstream region. For increasing
   back-reflections, the approximate result (\ref{260}) overestimates the
   ``true'' value of the transmission coefficient given by Eq.~(\ref{330}).
}
\end{figure}

For the practical computation of the transmission value associated with a given
scattering state, it is sufficient to evaluate the integral (\ref{290})
numerically in the near-upstream region (i.e.\ for $x \simeq x_2$ in
Fig.~\ref{Fig_Adpass}) where the extremal densities $n_\pm$ can be found by
solving $W(n_{\pm})=E_u$.
This means that the adiabatic variation of $g$ does not need to be included
at all in the calculation; it is sufficient to take into
account a short spatial domain in the upstream region within which the
condensate exhibits a couple of density oscillations.
Computing the transmission by use of Eq.~(\ref{330}) circumvents the
approximate character of the relation (\ref{260}) and is therefore also valid
in the regime of strong atom-atom interactions as well as for large
back-reflections.  
In Fig. \ref{Fig_Transvergleich} we compare the approximate with the ``exact''
expression for the transmission, determined by Eqs.~(\ref{260}) and
(\ref{330}) respectively, for a condensate with a moderate nonlinearity that
encounters a potential barrier in the guide.
For small large back-reflections both results coincide, whereas for large
back-reflections the approximate formula (\ref{260}) systematically
overestimates the transmission.

\subsection{Time-dependent transport processes}\label{sec2C}

So far, we restricted our considerations to stationary scattering solutions of
the time-independent Gross-Pitaevskii equation.
A severe problem is the fact that the mere existence of a stationary
scattering state does not imply that this state is dynamically stable and can
be populated in a time-dependent scattering process.
This is not only true for the propagation of a finite wave packet (which
obviously  cannot be evolved by an expansion in terms of stationary solutions
of the \GP, due to the absence of superposition principle), but also concerns
the limiting case of a quasi-stationary flow that is generated by an adiabatic
injection of the condensate into the waveguide.
This affects, as we shall discuss later on, the resonant transport of a
condensate through a double barrier potential, where the dynamical stability
properties of the scattering states become crucial for their population.

\begin{figure}[ptb]
\centering
\includegraphics[width=1.0\linewidth,angle=0]{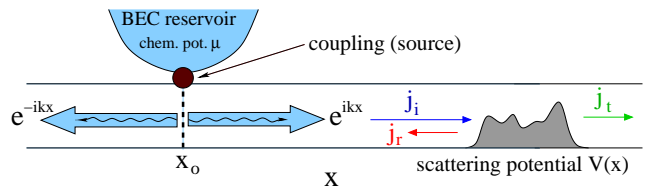}
\\[4mm]
 \caption{\label{Fig_Ankopplung} (color online) A reservoir of Bose-condensed
   matter with a given chemical potential $\mu$ is locally coupled at the
   position $x_0$ to a waveguide with a scattering potential. The reservoir
   emits a plane matter wave in both directions into the guide. Hence, a
   coherent beam, with current $j_i$, propagates towards the barriers of the
   potential where the condensate is partially reflected, with the current 
   $j_r$, and partially transmitted, with the current $j_t$.
}
\end{figure}

In view of this complication we now describe a method based on the 
{\em time-dependent} \GP, which allows us to simulate a realistic propagation
process. This equation is integrated in presence of an inhomogeneous
{\em source term}, located at a position $x=x_0$ in the upstream region and
emitting monochromatic matter waves. The source term simulates the coupling of
the waveguide to a large reservoir of a Bose-condensed matter at a given
chemical potential $\mu$, from which matter waves are injected into the
waveguide (see Fig.~\ref{Fig_Ankopplung}).
The effective nonlinear wave equation that governs the time evolution of the
condensate wave function $\psi(x,t)$ is therefore given by
\begin{eqnarray}\label{340}
 i \hbar \frac{\partial\psi(x,t) }{\partial t}&=& \left[ -\frac{\hbar^2}{2m}
 \frac{\partial^2 }{\partial x^2} + V_{||}(x) +g\vert\psi(x,t)\vert^2 \right]
\psi(x,t),
 \nonumber\\[3mm]
 &&+S(t)~\delta(x-x_{0})~\exp{(-i \mu t /\hbar)}, \; 
\end{eqnarray}
where the time-dependent coupling strength between the waveguide and the
reservoir is contained within the source amplitude $S(t)$.
The interaction parameter $g$ need not be constant, but may be considered to
position-dependent as well, in order, e.g., to simulate the adiabatic
transition from a noninteracting to an interacting guide as depicted in
Fig.~\ref{Fig_Adpass}.
In this work, we restrict ourselves to the case where $g$ is 
constant in the vicinity of the finite-range scattering potential.

Before studying time-dependent scattering processes in a waveguide with a
finite scattering potential, it is instructive to consider first stationary
solutions of Eq.~(\ref{340}) for the particular case of a homogeneous
waveguide, i.e.\ $V_{||}(x)\equiv 0$, and a constant source amplitude $S(t)\equiv S_0$.
In this case, there exist plane wave solutions $\psi(x,t)=\psi(x)e^{-i\mu t}$ with
constant density $n=\vert \psi(x,t) \vert^2 $.
To demonstrate this, we switch to the Fourier space by introducing the Fourier
transformed wave function $\tilde \psi(q,t)=\int \exp(iqx)\psi(x,t)dq$.
Then, Eq.~(\ref{340}) takes the form
\begin{equation} \label{350}
\left({i}{\hbar}\frac{\partial}{\partial t} -\frac{\hbar^2 \, q^2}
{2 m} -g n \right)\tilde\psi(q,t)=S_{0}\,e^{-i q x_0}\,e^{-i\mu  t/ \hbar }
\; .
\end{equation}
This equation admits solutions of the form
\begin{equation} \label{360}
\tilde\psi(q,t)=\frac{2 m S_{0}\,e^{-i q x_0}}
{\hbar^2 k^2 -\hbar^2 q^2}\,e^{-i\mu t/\hbar}
 \; .
\end{equation}
Here, we introduced the wave vector $k$ via the relation $\hbar^2 k^2 = 2m(\mu-gn)$.
By transforming back to the position space, we find solutions where the source
term emits in both directions the monochromatic wave
\begin{equation} \label{370}
\psi(x,t)=\frac{S_{0} m}{i k \hbar^2}\;e^{i k \vert x-x_{0}\vert
}\;e^{-i \mu t/\hbar} ,
\end{equation}
with the wave number $k$ being self-consistently defined by
\begin{equation} \label{380}
k^2= \frac{1}{\hbar^2}\left[{2m \left(\mu  -  g \frac{|S_0|^2m^2}{\hbar^4 k^2}  \right)}\right].
\end{equation}
The density that is associated with the wave function (\ref{370}) reads
\begin{equation} \label{390}
n=\frac{|S_0^2|m^2}{( \hbar^4 k^2)}=\frac{|S_0|^2 m}{2\hbar^2(\mu - gn)}.
\end{equation}
Evaluating the quantum mechanical current operator shows that the source emits
the current 
\begin{equation} \label{400}
j_i =  \pm \frac{| S_0 |^2 m}{(\hbar^3 k)}\ \overset{(\ref{390})}{ = } \ \pm
\frac{1}{\hbar} |S_0|\sqrt{n}, 
\end{equation}
with `` $+$ `` for $x>x_0$ and ``$-$ `` for $x<x_0$.
Inserting Eq.~(\ref{400}) into  Eq.~(\ref{390}) immediately yields the 
plane-wave dispersion relation $\mu=m j_i^2/(2 n^2) + gn$.

\begin{figure}[ptb]
\centering
\includegraphics[width=0.9\linewidth,angle=0]{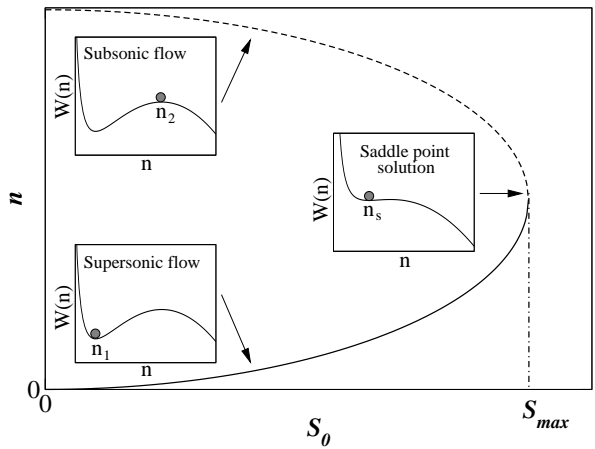}
\\[4mm]
 \caption{\label{Fig_illu} Illustration of the relation (\ref{410})
   between the source amplitude $S_0$ and the densities $n_{1,2}$ of a
   homogeneous flow. The lower branch corresponds to the supersonic solutions
   with density $n_1$ and the upper branch to the subsonic solutions with
   density $n_2$. The insets illustrate the configuration of the classical
   potential $W(n)$ for the different types of solutions.
}
\end{figure}

As already discussed in Sec.~\ref{sec2B} (see Eq.~(\ref{180})) this equation
admits two flat-density solutions, $n=n_1$ and $n=n_2$, corresponding to a
supersonic and a subsonic propagation of the condensate, respectively.
Rewriting Eq.~(\ref{390}) in the form
\begin{equation} \label{410}
n_{1,2}=\frac{1}{2 g}\left(\mu \mp \sqrt{\mu^2-2gm |S_0|^2 / \hbar^2}  \right)
\end{equation}
allows one to compute the two densities $n_{1,2}$ that are possible for a
given value of the source amplitude $S_0$.
This relation is illustrated in Fig.~\ref{Fig_illu}: the lower branch contains
the supersonic solutions and the upper branch the subsonic solutions. 
The value $S_{max}=\hbar \mu / \sqrt{2gm}$ corresponds to the saddle point
configuration of the classical potential $W(n)$; for source amplitudes larger
than this threshold, no stationary solutions of Eq.~(\ref{340}) are possible.
In the limit of noninteracting particles, $g=0$, only the supersonic branch
survives (because the speed of sound is zero) and Eq.~(\ref{410}) takes the
simple form $n=|S_0|^2 m /(2\hbar^2\mu)$.

Now we study the time evolution of $\psi(x,t)$ in presence of a variable source
amplitude $S(t)$. Here, the scenario of an initially empty waveguide
that is gradually filled with matter waves is of peculiar interest as
this corresponds to the experimentally realistic situation where the
condensate is initially confined in a microtrap (playing the role of the
reservoir) and then smoothly released to propagate into the waveguide.
To simulate such a process, we propagate $\psi(x,t)$ by numerically integrating
the wave equation (\ref{340}) in presence of an adiabatic increase of the
source amplitude $S(t)$ from $S(t=0)=0$ up to a given maximal value $S_0$, with
the initial condition $\psi(x,t=0) \equiv 0$. 
The amplitude $S(t)$ is increased adiabatically in order to ensure that, at
any instant during the propagation, the wave function in the guide remains as
close as possible to a stationary scattering state of the form 
$\psi(x)\exp(-i \mu t /\hbar)$.
Quantitatively this means that the typical time scale $\Delta T$ on which the
amplitude $S(t)$ increases is much larger than the characteristic time scale 
$\tau \equiv \hbar / \mu$ that is associated with the chemical potential $\mu$ of the source:
$\Delta T\gg \tau$.
As we are studying an infinitely extended scattering problem, we have to impose
absorbing boundary conditions in order to avoid artificial back-reflection at
the boundaries of the numerical grid. 
Details on these absorbing boundaries, which are taken from
Ref.~\cite{Shibata} and adapted to account also for a finite nonlinearity,
as well as on the numerical integration procedure are given in Appendix
\ref{Appendix2}.

\begin{figure}[ptb]
\centering
\includegraphics[width=0.8\linewidth,angle=0]{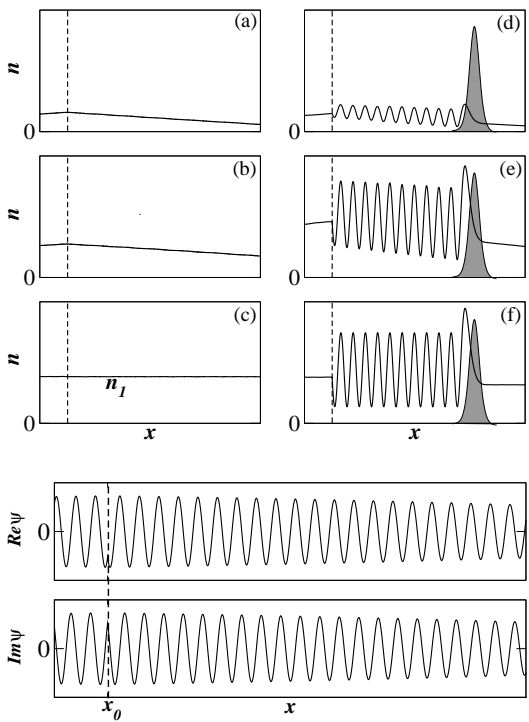}
\\[5mm]
\includegraphics[width=0.8\linewidth,angle=0]{Quelltest1.eps}   
\\[4mm]
 \caption{\label{Fig_einschalt} Time evolution of the condensate density
   during the adiabatic increase of the source amplitude (the source is
   located at the vertical dashed lines).
   The panels (a-c) show three snapshots of the condensate in a waveguide
   without scattering potential: at (a) $t=0.1\,\Delta T$,  (b) $t=\Delta T$, and (c)
   $t=10\,\Delta T$.
The bottom part of the figure shows the real and imaginary part of the wave
function whose density is displayed in panel (b). The panels (d-f) illustrate
the scattering of the matter waves at a repulsive barrier potential 
(gray-shaded region). Panel (f) clearly shows that a stationary scattering
state is populated in the long-time limit $t\gg \Delta T$.
}
\end{figure}

In a first step we discuss the filling of the waveguide in absence of a
scattering potential, i.e. for $V_{||}(x) \equiv 0$.
Fig.~\ref{Fig_einschalt}(a-c) displays the time-evolution of the wave function
$\psi(x,t)$ by a series of snapshots showing the density at different times. For
the sake of definiteness we chose $S(t)=S_0 [1-\exp(-t/\Delta T)]\ $, which
provides a smooth evolution towards the desired final value $S(t\to\infty)=S_0$.
We find that for propagation times $t \gg \Delta T$ the calculation converges towards 
the flat density (Fig.~\ref{Fig_einschalt}c) that corresponds to the
stationary plane wave (\ref{370}) at the source amplitude $S=S_0$.
The bottom part of Fig.~\ref{Fig_einschalt} shows the real and imaginary parts
of the wavefunction $\psi$ at the time $t=\Delta T$ during the filling 
process.
These panels clearly illustrates that the source emits a plane-wave like
solution of the form $A(x,t)\exp[-i\mu t+ik(x,t)x]$ where $A(x,t)$ and $k(x,t)$
vary slowly with position and time.

\begin{figure}[t]
\centering
\includegraphics[width=0.9\linewidth,angle=0]{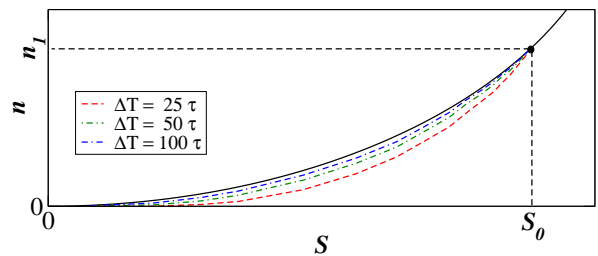}
\caption{\label{Fig_Adiabaten}
  (color online) Evolution of the condensate density as a function of the
  source amplitude $S$ for three different values of the time scale $\Delta T$ in
  which $S$ is ramped to its maximal value $S_0$ (dashed and dashed-dotted
  curves). For an increasing ratio $\Delta T / \tau$ with $\tau \equiv \hbar / \mu$, the curves
  converge towards the supersonic branch of Eq.~(\ref{410}) (solid line).
  For $S\to S_0$ the supersonic scattering state with constant density $n=n_1$
  is reached.
}
\end{figure}

It is instructive to display the evolution of the condensate density as a
function of the time-dependent source amplitude $S$. 
Fig.~\ref{Fig_Adiabaten} shows this evolution for different values of $\Delta T$
(dashed lines). We notice that these curves approach the lower branch
of the relation (\ref{410}) (solid line in Fig.~\ref{Fig_Adiabaten}, see also
Fig.~\ref{Fig_illu}) if we reach the limit $\Delta T\gg \tau$.
We therefore deduce that the adiabatic filling of an initially condensate-free
waveguide can only populate stationary solutions that correspond to a
supersonic flow; hence, the final condensate density is given by $n=n_1$ as
defined by Eq.~(\ref{410}).
In analogy to the fixed output problem discussed in Sec.~\ref{sec2B}, the
implementation of the source term therefore allows one to investigate the
transport of the condensate in terms of a so-called {\em fixed input problem},
where the incident current $j_i$ that is emitted into the guide parametrizes
the process. The fixed input approach is much closer to
experimental situations because the current that is injected into the guide is
typically under much better control than the total transmitted current
$j_t$.

In a second step, we consider a scattering process in presence of a barrier
potential $V_{||}(x)$.
Due to the partial backscattering of the condensate at the barrier, the
dynamics becomes more complex as compared to the potential-free case.
Nevertheless, for weak or moderate nonlinearities the wave function $\psi(x,t)$
is found to converge towards a stationary scattering state $\psi(x)\exp(-i \mu t
/\hbar)$
during the adiabatic increase of the source amplitude towards its final value
$S_0$, as illustrated in Fig.~\ref{Fig_einschalt}(d - f). During the gradual
filling of the guide, the condensate is partially reflected at the barrier,
which leads to the oscillating density pattern between the barrier and the
position of the source in the upstream region. On the right-hand side of the
barrier, in the downstream region, the density is flat in the long-time
limit $t\gg \Delta T$, which reflects the fact that the wave function $\psi(x)$ is given
there by an outgoing plane wave of the form $\psi(x)=A \exp(ikx)$.
We checked that the state $\psi(x)$ that is reached at the end of the propagation
fulfills the stationary Gross-Pitaevskii equation, i.e., the wave function's 
amplitude $A(x)={|\psi(x)|}$ is a solution of Eq.~(\ref{160}).

Once we populate a stationary state, we have another straightforward access to
the transmission coefficient $T$ in the nonlinear scattering problem: $T$ is
given by the ratio of the transmitted current $j_t$, evaluated through the
current operator in the downstream region, to the current that would
propagate through the waveguide in {\em absence} of the barrier potential,
which is the current $j_i$ that is directly emitted from the source. This
approach provides another natural extension of the definition of transmission
coefficients to nonlinear wave equations.
Hence, the numerical method introduced in this section allows not only to
calculate scattering states that are dynamically stable and can be populated
in a realistic propagation process \cite{Bemerkung1}, but provides also a
straightforward access to transmission coefficients for a fixed input problem.

We point out that in the nonlinear case, convergence towards a stationary
scattering state is not always guaranteed. Indeed, studying the transport of
condensates through a waveguide with an extended disorder region by means of
the method described in this section revealed that, beyond a critical
interaction strength respectively a critical length of the disorder region,
the transport process generally remains time-dependent and stationary states
are not populated \cite{PauO05PRA}.

\subsection{Transport through a quantum point contact}\label{sec2D}

As a first and simple example we study the transport of a Bose-Einstein
condensate through a quantum point contact. We consider a waveguide with a
constriction given by a single repulsive Gaussian barrier potential
$V_{||}(x)=V_0\exp(-x^2/\sigma^2)$ which can be experimentally implemented  by
focusing a blue detuned laser beam in its transverse ground mode onto the
waveguide \cite{Engels}.
For the sake of definiteness we consider in the following a condensate of
$^{87}$Rb atoms ($m=1.45\times 10^{-25}kg$, $a_s=5.77$ nm) flowing through a
waveguide with transverse trapping frequency $\omega = 2\pi \times 10^3$ s$^{-1}$ that
corresponds to a harmonic oscillator length $a_{\perp}=0.34$ $\mu$m. It is convenient
to measure energies in units of $\hbar \omega$, lengths in units of $a_{\perp}$ and
particle currents in units of $\omega$. In these units the interaction parameter
reads $g=0.034\hbar\omega a_{\perp}$. For the longitudinal extension of the barrier we
assume $\sigma = 2 a_{\perp} \simeq 0.7 \mu m$ (which would be at the limit of experimental
realizability) and its height is chosen as $V_0=3\,\hbar\omega$.

\begin{figure}[ptb]
\centering
\includegraphics[width=0.9\linewidth,angle=0]{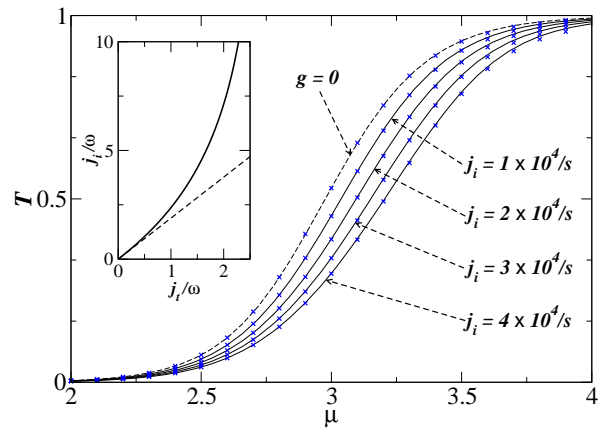}
\\[4mm]
 \caption{\label{Fig_QPTransm} (color online) Transmission spectrum of the
   condensate flow through a quantum point contact for different values of the
   incident current $j_i$ ($\mu$ in units of $\hbar\omega$). The solid lines are found by
   evaluating the stationary Gross-Pitaevskii equation, the values denoted by
   crosses are obtained by integrating the time-dependent Gross-Pitaevskii
   equation in presence of the source term (the dashed line displays the
   result for a noninteracting condensate).
The inset shows the $j_t - j_i$ current characteristics for a condensate flow
without interactions ($g=0$, dashed line) and in presence of interactions
($g=0.034\hbar\omega$, solid line), at $\mu=3\hbar\omega$.
}
\end{figure}

In a first step, we investigate the transport process in terms of a fixed
output problem: We calculate scattering states by integrating the
stationary Gross-Pitaevskii equation,
\begin{eqnarray} \label{415}
\mu \psi(x) = \left[-\frac{\hbar^2}{2 m} \frac{\partial^2}{\partial x^2}+V_0 e^{-x^2/\sigma^2}
  +g|\psi(x)|^2\right]\psi(x),
\end{eqnarray}
for given values of the chemical potential $\mu$ and the transmitted current
$j_t$ from the downstream to the upstream region (where a supersonic density
$n(x\to \infty)=n_1$ is assumed in the downstream region). 
Eq.~(\ref{330}) allows us to compute the corresponding transmission
coefficient from which we can deduce the incident current via $j_i=j_t / T$.
Varying the transmitted current allows us to compute the $j_t - j_i$ current
characteristics which is displayed for $\mu=3\hbar\omega $ in the inset of
Fig.~\ref{Fig_QPTransm}. For noninteracting particles ($g=0$) the $j_t - j_i$
characteristics is linear, because the transmission coefficient $T$ does not
depend on the particle current, whereas for non-vanishing interaction
parameters the $j_t - j_i$ characteristics shows a nonlinear behavior and
displays increasing deviations from the linear case with increasing particle
currents. This means that the presence of repulsive interactions suppresses
the transmission through the quantum point contact with growing current.

It is now easy to switch from the fixed output to the fixed input problem
where the incident current $j_i$ is kept constant. 
To this end, we basically have to invert the $j_t - j_i$ characteristics, in
order to determine the total current $j_t$ and the corresponding transmission
coefficient $T$ that result from a given incident current $j_i$.
This can be done in a unique way in the present case, since the $j_t - j_i$
characteristics is monotonous and therefore allows one to unambiguously
assign to each value of  $j_t$ a unique incident current $j_i$.
Computing the current characteristics for different values of $\mu$ allows one
then to obtain the transmission spectrum, i.e.\ the transmission coefficient
$T$ as a function of the chemical potential $\mu$ at a fixed incident current
$j_i$. 
Transmission spectra of the point contact for different incident currents
are displayed in Fig.~\ref{Fig_QPTransm}.
Qualitatively, we find that in presence of repulsive interactions the spectra
resemble strongly the spectrum for a single particle: 
for chemical potentials considerably smaller than $V_0$ the transmission tends
to zero, whereas for $\mu$ much larger than $V_0$ we reach a regime of perfect
transmission. In the intermediate regime, we clearly see that
increasing particle currents $j_i$ yield a moderate suppression of the
condensate flow through the point contact. This is attributed to the fact that
the presence of the repulsive interaction leads, at fixed $\mu$, to a reduction
of the available kinetic energy, which in turn reduces the probability for
the atoms to penetrate the barrier.

So far, the computation of the transmission spectra was based on the
stationary Gross-Pitaevskii equation. 
As a complementary access, we apply the method based on integrating the
time-dependent Gross-Pitaevskii equation with source term. For each value of
$\mu$, the wave function was propagated according to Eq.~(\ref{350}) in the
presence of an adiabatic increase of the source amplitude $S$ up to the
maximum value $S_0$ that corresponds to a given incident current $j_i$. For
the considered range of incident currents $j_i$ we find stationary scattering
states at the end of the propagation.
As shown in Fig.~\ref{Fig_QPTransm}, the results for the transmission obtained
from the time-dependent integration (marked by blue crosses) coincide with the
result based on evaluating the stationary Gross-Pitaevskii equation. Hence we
can conclude that a gradual filling of the guide populates precisely those
scattering states that are eigenmodes of the stationary problem, and that
these stationary states are dynamically stable. 

\section{Transport through a double barrier potential}\label{sec3}

Now we study the particularly interesting propagation process of a
Bose-Einstein condensate through a symmetric repulsive double barrier
potential which can be seen as a Fabry-Perot interferometer for matter waves. 
This setup was first discussed by Carusotto and La Rocca
\cite{CarLar99PRL,Car01PRA} who proposed to use a combination of optical
lattices for the realization of this bosonic quantum dot. 
In the context of atom chips, a double barrier potential could also be
implemented by suitable geometries of microfabricated wires on a multilayer
chip geometry \cite{Bem1}. 
Another straightforward implementation relies on two blue-detuned parallel
laser beams, crossing transversely the waveguide. Assuming the laser beams to
be in the lowest transverse mode, this setup creates a potential geometry with
two Gaussian shaped barriers. 
For the sake of definiteness, we consider this latter case and assume a double
barrier given by 
\begin{equation}\label{DB10}
V_{db}(x)= V_0\left[e^{-(x+L/2)^2 / \sigma ^2} +  e^{-(x-L/2)^2 /  \sigma^2}  \right].
\end{equation}
Here, $\sigma$ is the width of one barrier and $L$ is the distance between the
barriers. 

For a flow of noninteracting particles it is well known that the transmission
spectrum of a symmetric double barrier potential exhibits
{\em Breit-Wigner resonances} \cite{Ferry} which are related to resonant
transport states.
In our context, these resonant states can be defined as stationary scattering
states of the condensate (see Eq.~(\ref{186})$\,$) that exhibit perfect
transmission.
In the following, we investigate to which extent resonant transport through
such a double barrier potential can be achieved for an interacting condensate,
and how interactions modify the transmission spectrum. 

\subsection{Resonant transmission spectra}\label{sec3A}

\begin{figure}[ptb]
\centering
\includegraphics[width=1.0\linewidth,angle=0]{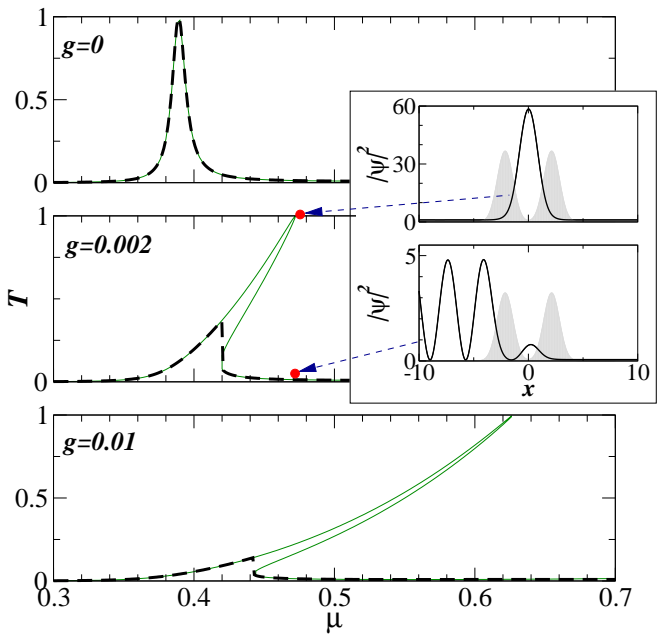}
\\[4mm]
 \caption{\label{Fig_ResonanzSpektrum} (color online) Transmission spectra of
   the double barrier potential  at $g = 0$ (upper panel), $g = 0.002$ $\hbar \omega
   a_{\perp}$ (middle panel), and $g =0.01$ $\hbar \omega a_{\perp}$ (lower panel). 
The (green) solid lines show the transmissions of all scattering states,
calculated by the ``stationary'' method based on Eq.~(\ref{160}), that exist
at the incident current $j_i=1.0\, \omega$ of the matter-wave beam. 
The dashed lines display the spectra obtained from the time-dependent
integration approach. 
The inset shows the longitudinal atom densities (in units of $a_{\perp}^{-1}$) of
the first resonant state and the coexisting low-transmission state for $g =
0.002$ $\hbar \omega a_{\perp} $ (the position $x$ is given in units of $a_{\perp}$). The
gray-shaded curves indicate the positions of the two barriers. 
The (red) dots (marked by the arrows) designate the positions of the resonant
state and the low-transmission state in the transmission spectrum. 
}
\end{figure}

We now compute transmission spectra for the double barrier potential
(\ref{DB10}) by applying the same methods that have been employed to find the
spectra of the quantum point contact in Sec.~\ref{sec2D}, using again the same
units that were already introduced there. In the following, 
we consider a condensate with effective interaction strength $g$
(which will be varied to investigate the effect of an increasing
nonlinearity), a waveguide with transverse trapping frequency 
$\omega = 2\pi\times 10^3s^{-1}$, and a double barrier potential (\ref{DB10}) with the
parameters $V_0=1.1\,\hbar\omega$, $\sigma = a_{\perp}$, and $L=4.25 a_{\perp} $.
We study the transport of the condensate in terms of a fixed input problem,
with incident current $j_i = 1.0\, \omega$.
The influence of the atom-atom interaction on the transmission spectrum is
exemplarily investigated in the vicinity of the energetically lowest resonance
which has one density maximum in between the two barriers (see inset in
Fig.~\ref{Fig_ResonanzSpektrum}). 
In Ref.~\cite{PauRicSch05PRL} we showed that qualitatively similar results are
also found for higher resonances.

First we compute transmission spectra by use of the integration method based
on the stationary Gross-Piatevskii equation, respectively Eq.~(\ref{160}):
The spectrum is, as in Sec.~\ref{sec2D}, determined by calculating stationary
scattering states for given $j_t$ and $\mu$, and the incident current of the
scattering states is computed via Eq.~(\ref{330}).
Finding the value of $j_t$ that results from a given $j_i$ is an optimization
problem that can be solved systematically by analyzing the $j_t-j_i$ current
characteristics.
In the linear case, $g=0$, we obtain a Breit-Wigner resonance at $\mu=0.389 \hbar \omega
$ corresponding to  the energetically lowest resonance state
(Fig.~\ref{Fig_ResonanzSpektrum}).

Now we consider the case of a weak atom-atom interaction, $g=0.002\hbar\omega
a_{\perp}$. As the most striking result, we find, close to the resonance, a
multivalued transmission spectrum where two further solutions appear for
$0.419< \mu /(\hbar\omega) <0.472$. These solutions join together to form a resonance
peak that is asymmetrically distorted towards higher values of the chemical
potential \cite{Bem333}. 
The resonant state, which is found at $\mu=0.472\hbar\omega$, coexists with a
low-transmission state, as depicted in the central panel of
Fig.~\ref{Fig_ResonanzSpektrum}.
The asymmetric distortion becomes even more pronounced for an increasing
interaction strength. This is shown  in the bottom panel of
Fig.~\ref{Fig_ResonanzSpektrum} where we dipslay the spectrum in the vicinity
of the first resonance for $g=0.01\hbar\omega_{\perp}$.  

\begin{figure}[ptb]
\centering
\includegraphics[width=0.98\linewidth,angle=0]{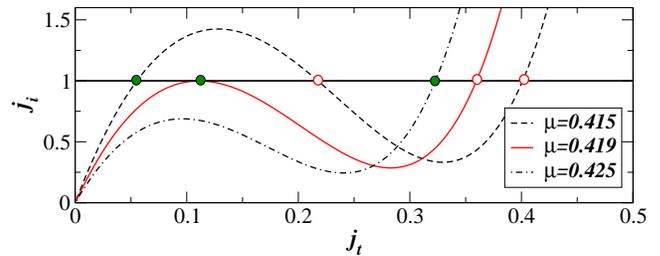}\\[4mm]
 \caption{\label{Fig_stromcharacteristic} (color online) Current
   characteristics for the double barrier potential in the vicinity of the
   onset of the multivalued subzone of the spectrum (for $g = 0.002$ $\hbar \omega
   a_{\perp} $). Below the critical $\mu=0.419\hbar\omega$, the $j_i-j_t$ characteristics
   (dashed-dotted line) intersects only once the horizontal line indicating
   the fixed incident current $j_i = 1.0\, \omega$. Above this critical value,
   three intersection points are found (dashed line). At $\mu=0.419\hbar\omega$, the
   current characteristics (solid line) exhibits a tangent to the horizontal
   line. The intersection points marked with filled (green) points correspond
   to scattering states that are populated during a time-dependent propagation
   process. 
}
\end{figure}

It is instructive to trace the evolution of the $j_t-j_i$ characteristics in
the vicinity of the onset of the multivalued subzone in the spectrum. 
In contrast to the monotonously increasing current characteristics that we
found for the quantum point contact (Fig.~\ref{Fig_QPTransm}), the
characteristics of the double barrier potential shows a more complex behavior,
where it is not always possible to unambiguously attribute to each incident
current $j_i$ one single transmitted current $j_t$.
Fig.~\ref{Fig_stromcharacteristic} shows that for values of $\mu$ below the
critical chemical potential from which on three branches coexist, the current
characteristics intersects only once the horizontal line that represents the
fixed incident current $j_i = 1.0\, \omega$. Above this critical value of $\mu$ three
intersection points are found, corresponding to the three coexisting
scattering states. 

Our findings are characteristic for a bistability phenomenon,  similar to
processes in nonlinear optics \cite{BoydBook} and in the electronic transport
through quantum wells \cite{Goldman87,Azbel99}. It is crucial to know which
branches of the transmission spectrum are actually populated in a realistic
experimental situation in order to decide if resonant transport is possible in
presence of a finite interaction strength. 
To this end, we recalculate the transmission spectrum with the
\emph{time-dependent} integration approach, which simulates, at given value of
$\mu$, the adiabatic release of the condensate from the reservoir into the
waveguide.
As explained in Sec.~\ref{sec2D}, this method provides another straightforward
access to the transmission values, and stationary states that are selected by
this method automatically satisfy the criterion that they are dynamically
stable and can be populated in a realistic propagation process. 

The dashed lines in Fig.~\ref{Fig_ResonanzSpektrum} show the result of this
calculation.
While a perfect agreement with the method based on the stationary
Gross-Pitaevskii equation is found for $g=0$, the time-dependent approach
reproduces, for $g \neq 0$, only the lowest branches of the spectra in the
multivalued region.
This apparently implies that the asymmetrically distorted peak structure
is essentially inaccessible in the propagation process that is considered
here.
We therefore conclude that resonant transport, which would necessarily
require the population of such a distorted peak, will generally be suppressed
in presence of finite interactions, and only the low branches of the
spectrum which have rather low transmission will be populated.
Qualitatively, this behavior of the nonlinear system can be understood by
comparing the ``internal'' interaction energy (evaluated within the internal
region of the double barrier)
\begin{equation}
E_{\text{int}}=g \int_{-L/2}^{+L/2} |\psi(x)|^2 \,dx
\end{equation}
of the resonant with the one of the coexisting low-transmission state.
The system can minimize $E_{\text{int}}$ by realizing a state with a low
particle density in between the barriers.
As displayed in the inset of Fig.~\ref{Fig_ResonanzSpektrum}, this favors the
low-transmission state.

To conclude this section, we remark that a temporary enhancement of the
transmission of matter waves near the resonance can be achieved by a variation
of the external potential during the propagation process. In
Ref.~\cite{PauRicSch05PRL} we devised a temporal modulation scheme where the
potential is shifted with time according to $V(x)\to V(x,t)\equiv V(x)-V_0(t)$.
Specifically, such a modulation can be induced by illuminating the scattering
region with a red-detuned laser pulse, where $V_0(t)>0$ would be determined by
the detuning and the intensity of the laser. 
In the case of an adiabatic modulation of $V$, the wave function $\psi(x,t)$
remains, at each time $t$, close to the instantaneous scattering state that is
associated with the external potential $V(x,t)$ --- or, equivalently
formulated, close to the scattering state for the potential $V(x)$ at the
shifted chemical potential $\mu + V_0(t)$.
As soon as $\mu + V_0(t)$ is raised  above the critical chemical potential from
which on the transmission spectrum becomes multivalued, the wave function
follows continuously the upper branch of the resonance and evolves into a
near-resonant scattering state with high transmission.
This state turns out to be dynamical unstable, and the wave function decays
after a typical lifetime of the order of several milliseconds 
towards a low-transmission state \cite{PauRicSch05PRL}.

\subsection{Transmission in terms of quasi-bound states}\label{sec3C}
In this subsection, we present analytical and numerical evidence that the
distortion of the resonance peak arises indeed due to the nonlinearity-induced
level shift of the self-consistent quasi-bound state within the atomic quantum
dot.
We describe, for this purpose, our system in a similar way as in the
well-known scattering matrix approach \cite{MahWei}, namely by a discrete
``bound'' (or quasi-bound) state within the quantum dot that is weakly coupled
to two symmetric continua of unbound ``lead'' states in the up- and downstream
regions of the waveguide.
In contrast to the situations for which the scattering matrix formalism was
originally developed \cite{MahWei}, we consider here \emph{nonlinear}
dynamics within the quantum dot, which is described by the
Gross-Pitaevskii equation.
As was pointed out above
the outcome of a given scattering process
is, in this case, not completely independent of the ``history'' of the
process, i.e., of the way in which the condensate is injected into the
waveguide.
Different scattering states might, specifically, be populated if the chemical
potential is adiabatically varied in different ways during the propagation
\cite{PauRicSch05PRL}.
To account for this complication, we formulate our nonlinear scattering theory
in a \emph{time-dependent} way, namely by considering the asymptotic
propagation of a spatially broad (and energetically narrow) wave packet that
is injected onto the quantum dot from the left (upstream) lead.
The population of the wave packet that exits the scattering region in the
right lead gives naturally rise to the transmission coefficient.

As starting point, we subdivde of the Hilbert space
$\mathcal{H}$ into a subspace $\mathcal{H}_0$ containing discrete bound states
within the quantum dot region, and two other subspaces $\mathcal{H}_{L/R}$
containing continuous states in the left and right leads of the waveguide.
This subdivision can be formally achieved by means of the 
\emph{Feshbach projection method} \cite{Fes58AP}, where those subspaces are
defined by the projection operators $P_L = \theta(x_L - \hat{x})$, 
$P_R = \theta(\hat{x} - x_R)$, and $Q = 1 - P_L - P_R$.
Here $x_L$ and $x_R$ are suitably chosen positions that mark the left and
right boundaries of the quantum dot, and $\theta$ denotes the Heavyside step
function.
As an essential ingredient of the Feshbach formalism, different boundary
conditions (i.e., of Dirichlet or Neumann type) are imposed within and outside
the dot, which allows one then to shift the boundary contributions from matrix
elements of the Laplace operator to appropriate sides of the spatial cuts at
$x = x_{L/R}$, in such a way that the operator $T$ of the kinetic energy
remains Hermitean within each subspace, but exhibits finite coupling matrix
elements across the boundaries (see, e.g., Ref.~\cite{VivHac03PRA} for more
details).
Choosing Dirichlet boundary conditions within the resonator and Neumann
boundary conditions in the leads, these matrix elements would read
\begin{eqnarray}
  \langle \psi_R | T | \phi \rangle & = & \frac{\hbar^2}{2m} \psi_R^*(x_R) \phi'(x_R) \label{eq:T0R} \\
  \langle \psi_L | T | \phi \rangle & = & - \frac{\hbar^2}{2m} \psi_L^*(x_L) \phi'(x_L) \label{eq:T0L}
\end{eqnarray}
for wave functions $\phi(x)$, $\psi_L(x)$, and $\psi_R(x)$ defined within the subspaces
$\mathcal{H}_0$, $\mathcal{H}_L$, and $\mathcal{H}_R$, respectively.
Without loss of generality, we set $x_L \equiv -a$ and $x_R \equiv a$ in the following,
where $a = L/2$ denotes the position of the maximal barrier height.

We now make the assumption that the nonlinearity can be neglected in the lead
regions outside the quantum dot, which should be valid at weak interaction
strengths and which is motivated by the fact that close to resonance the
density within the double barrier potential is strongly enhanced as compared
to the leads.
We furthermore assume that only one quasi-bound state, namely the local
``ground state'' of the quantum dot, appreciably contributes to the scattering
process, which is indeed the case in our specific double barrier potential
(\ref{DB10}) where ``excited'' quasi-bound states are energetically located
above the barrier height.
Neglecting the contribution of those excited states, we make the ansatz 
\begin{eqnarray}
  \psi(x,t) & = &  \int_0^\infty dE A_E^L(t) \phi_E^L(x) + B(t) \phi_0(x) \nonumber \\
  & & + \int_0^\infty dE A_E^R(t) \phi_E^R(x)
\end{eqnarray}
for the wave function, where $\phi_0 \in \mathcal{H}_0$ denotes the above
quasi-bound state and $\phi_E^{L/R} \in \mathcal{H}_{L/R}$ are the
energy-normalized continuum eigenstates within the left and right lead,
respectively, at energy $E$.
Inserting this ansatz into the Gross-Pitaevskii equation yields the
equations 
\begin{eqnarray}
  i \hbar \frac{d}{d t} A_E^{L/R}(t) & = & E A_E^{L/R}(t) + V_E B(t) \label{eq:At} \\
  i \hbar \frac{d}{d t} B(t) & = & \mu_0\left( |B(t)|^2 \right) B(t) \nonumber \\
  & & + \int_0^\infty dE \, V_E \left[ A_E^L(t) + A_E^R(t) \right] \label{eq:Bt}
\end{eqnarray}
for the amplitudes $A_E^L$, $A_E^R$, and $B$.
Here,
\begin{equation}
  \mu_0\left( |B(t)|^2 \right) \equiv \mu_0^{(0)} + \tilde{g} |B(t)|^2
\end{equation}
with
\begin{equation}
  \tilde{g} \equiv g \int_{-a}^a |\phi_0(x)|^2 dx 
\end{equation}
represents the population-dependent chemical potential of the quasi-bound
state, and
\begin{equation}
  V_E \equiv \frac{\hbar^2}{2m} \phi_0'(a) \phi_E^R(a) = - \frac{\hbar^2}{2m} \phi_0'(-a) \phi_E^L(-a) 
\end{equation}
denotes the coupling matrix element between $\phi_0$ and $\phi_E^{L/R}$.
We assume here, without loss of generality, that the wave functions $\phi_0(x)$,
$\phi_E^L(x)$, and $ \phi_E^R(x)$ are real-valued and that the continuum
eigenfunctions exhibit the symmetry-related property $\phi_E^R(x) = \phi_E^L(-x)$.

As appropriate initial state for the quasi-stationary scattering process, we
consider a spatially broad Gaussian wave packet that is injected from the
left-hand side onto the double barrier potential.
This wave packet is explicitly written as
\begin{equation}
  \psi(x,t_\epsilon) = \alpha \exp\left[ - \frac{(x + x_\epsilon)^2}{2 \sigma_\epsilon^2} + i k \left( x +
      \frac{1}{2} x_\epsilon \right) \right]
\end{equation}
with $x_\epsilon \equiv x_0 / \epsilon^3$ and $\sigma_\epsilon \equiv \sigma_0 / \epsilon^2$ for $x_0, \sigma_0 > 0$.
Choosing the initial time $t_\epsilon$ in the asymptotic past according to 
$t_\epsilon = - m x_\epsilon / (\hbar k)$, the wave packet will, in the limit $\epsilon \to 0_+$, evolve
into the plane wave
\begin{equation}
  \psi(x,t) = \alpha e^{i (k x - \mu t / \hbar )}
\end{equation}
at finite times $t$, with the incident chemical potential 
$\mu \equiv \hbar^2 k^2 / (2 m)$.
Using the fact that the energy-normalized continuum eigenfunctions are, in the
asymptotic spatial region $x \gg a$, given by
\begin{equation}
  \phi_E^R(x) = \phi_E^L(-x) = \sqrt{\frac{2m}{\pi \hbar^2 k_E}} \cos ( k_E x + \varphi_E )
\end{equation}
with $k_E \equiv \sqrt{2mE}/ \hbar$ and with a potential-dependent phase $\varphi_E$, we
obtain the initial amplitudes
\begin{eqnarray}
  A_E^L(t_\epsilon) & = & \sqrt{\frac{m \sigma_\epsilon^2}{\hbar^2 k_E}} \alpha \exp\left[ - \frac{1}{2} \sigma_\epsilon^2
    (k_E - k)^2 \right] \nonumber \\
  & & \times \exp \left[ + i x_\epsilon \left(k_E - \frac{k}{2}\right) + i \varphi_E \right]
\end{eqnarray}
and $B(t_\epsilon) = A_E^R(t_\epsilon) = 0$ for $\epsilon \to 0_+$.

Equation (\ref{eq:At}) can now be formally integrated yielding
\begin{eqnarray}
  A_E^{L/R}(t) & = & A_E^{L/R}(t_\epsilon) e^{- i E (t - t_\epsilon ) / \hbar} \nonumber \\
  & & - \frac{i}{\hbar} V_E \int_{t_\epsilon}^t B(t') e^{- i E (t - t' ) / \hbar } dt' \, .
  \label{eq:ALR}
\end{eqnarray}
Inserting this expression into Eq.~(\ref{eq:Bt}) leads to the equation
\begin{eqnarray}
  i \hbar \frac{d}{d t} B(t) & = & \mu_0\left( |B(t)|^2 \right) B(t) \nonumber \\
  & & - \frac{2 i}{\hbar} \int_{t_\epsilon}^t dt' \, B(t') e^{-i \mu(t - t') / \hbar} K(t - t') \nonumber \\
  & & + \int_0^\infty dE \, V_E A_E^{L}(t_\epsilon) e^{- i E (t - t_\epsilon ) / \hbar}
  \label{eq:Btmod}
\end{eqnarray}
for the bound component, with the Kernel
\begin{equation}
  K(\tau) = \int_0^\infty dE \, V_E^2 e^{-i(E - \mu) \tau / \hbar} \, .
\end{equation}
In the limit $\epsilon \to 0$, the last term on the right-hand side of
Eq.~(\ref{eq:Btmod}) is evaluated as $S e^{-i \mu t / \hbar}$ with the effective 
source amplitude
\begin{equation}
  S = \sqrt{\frac{2 \pi \hbar^2 k}{m}} V_\mu \alpha e^{i \varphi_\mu} \, . \label{eq:Source}
\end{equation}
This suggests that the time-dependence of the bound amplitude is, in the
quasi-stationary case, dominated by the exponential factor $e^{-i \mu t / \hbar}$.

This latter information permits now to evaluate the second term on the
right-hand side of Eq.~(\ref{eq:Btmod}):
if $B(t') \exp(i \mu t' / \hbar)$ varies much more slowly with time than 
$K(t - t')$, we can justify the approximation
\begin{eqnarray}
  \int_{t_\epsilon}^t dt' \, B(t') e^{-i \mu(t - t') / \hbar} K(t - t') & \simeq & B(t) \int_0^\infty d \tau
  K(\tau) \nonumber \\
  = \frac{ i \hbar}{2} \left( \delta_\mu - \frac{i}{2} \hbar \gamma_\mu \right) & &
\end{eqnarray}
where the energy shift $\delta_\mu$ and the rate $\gamma_\mu$ are, respectively, given by
the principal value integral
\begin{equation}
  \delta_\mu = \mathcal{P} \!\!\!\!\!\! \int dE \frac{2 V_E^2}{\mu - E} \, .
\end{equation}
and by the expression
\begin{equation}
  \gamma_\mu = 4 \pi V_\mu^2 / \hbar \, .
\end{equation}
Omitting the small shift $\delta_\mu$ in the following, we obtain the equation
\begin{eqnarray}
  i \hbar \frac{d}{d t} B(t) & = & \left( \mu_0\left( |B(t)|^2 \right) - \frac{i}{2}
    \hbar \gamma_\mu \right) B(t) \nonumber \\ 
  & & + S e^{-i \mu t / \hbar} \label{eq:BtS}
\end{eqnarray}
for the bound component $B(t)$, which exhibits strong analogies to a nonlinear
damped oscillator model that is subject to a periodic driving.
Obviously, stationary solutions of Eq.~(\ref{eq:BtS}) are of the form
\begin{equation}
  B(t) = B_0 e^{- i \mu t / \hbar}
\end{equation}
where the bound amplitude $B_0$ satisfies the self-consistent equation
\begin{equation}
  B_0 = \frac{S}{ \mu - \mu_0\left( |B_0|^2 \right) + \frac{i}{2} \hbar \gamma_\mu} \, .
  \label{eq:B0}
\end{equation}
For the noninteracting case $g = 0$, one can show that this solution is
necessarily realized after a transient propagation time of the order of
$\gamma_\mu^{-1}$.

Inserting this stationary solution into the equation (\ref{eq:ALR}) for the
transmitted component finally yields 
\begin{equation}
  A_\mu^R(t) = - 2 \pi i \frac{V_\mu^2 e^{-i \mu( t - t_\epsilon)/ \hbar}}{\mu - \mu_0\left( |B_0|^2 \right) +
    \frac{i}{2} \hbar \gamma_\mu} A_\mu^L(t_\epsilon)
\end{equation}
while $A_E^R(t)$ would, for $E \neq \mu$, vanish in the limit $\epsilon \to 0$.
We therefore obtain the transmission coefficient through
\begin{equation}
  T(\mu) \equiv \frac{|A_\mu^R(t)|^2}{|A_\mu^L(t_\epsilon)|^2} = 
  \frac{(\hbar \gamma_\mu / 2)^2}{\left[\mu - \mu_0\left( |B_0|^2 \right) \right]^2 + (\hbar \gamma_\mu
    / 2)^2} \, .
  \label{eq:trans}
\end{equation}
In the noninteracting limit $g \to 0$, this expression describes the
Breit-Wigner profile of a single resonance peak at $\mu = \mu_0$.
Indeed, if the decay rate $\gamma_\mu$ is sufficiently small around this resonance,
we can safely approximate $\gamma_\mu$ by $\gamma_{\mu_0}$ in the relevant energy range 
$|\mu - \mu_0| \lesssim \hbar \gamma_{\mu_0}$. Then $T(\mu)$ is given by a perfect
Lorentzian centered around $\mu = \mu_0$ with the width $\hbar \gamma_{\mu_0}$.
At finite $g \neq 0$, however, $T$ may exhibit several branches for a given value
of $\mu$, due to the implicit relation (\ref{eq:B0}) between the bound component
$B_0$ and the incident chemical potential $\mu$.

We now aim at reproducing the numerically calculated transmission
spectrum (see Fig.~\ref{Fig_ResonanzSpektrum}) through Eqs.~(\ref{eq:trans}) and (\ref{eq:B0}) using
information that is obtained from the corresponding \emph{decay} problem
\cite{MoiO04JPB,WitMosKor05JPA,CarHolMal05JPB,SchPau06PRA,WimSchMan06JPB},
namely the chemical potential and the instantaneous decay rate of the local
quasi-bound state at given population $|B_0|^2$.
The latter quantity can also be derived from Eq.~(\ref{eq:Btmod}), now in
\emph{absence} of the incident wave $A_E^{L}(t_\epsilon)$ and with the initial
population $B(t_0) = B_0$.
Taking into account the fact that the dominant time-dependence of $B(t)$ is,
in this case, given by $\exp[-i \mu(|B_0|^2) t / \hbar]$ for not too long evolution
times $t$, we obtain
\begin{equation}
  i \hbar \frac{d}{d t} B(t) = \left( \mu_0\left( |B(t)|^2 \right) - \frac{i}{2} \hbar
    \gamma_0\left( |B(t)|^2 \right) \right) B(t) \label{eq:Btd}
\end{equation}
as equation for the bound component $B(t)$, with 
$\gamma_0\left( |B|^2 \right) \equiv \gamma_{\mu_0\left( |B|^2 \right)}$.
Clearly, Eq.~(\ref{eq:Btd}) describes a \emph{nonexponential} decay of the
condensate in the quantum dot, which is explicitly given by the equation
\begin{equation}
  \frac{d}{dt} N_b(t) = - \gamma_0\left[ N_b(t) \right] N_b(t) \label{eq:dNdt}
\end{equation}
where the decay rate varies adiabatically with the remaining population
$N_b(t) \equiv |B(t)|^2$ of the quasi-bound state.
Such nonexponential decay processes of Bose-Einstein condensates were
discussed in detail in
Refs.~\cite{CarHolMal05JPB,SchPau06PRA,WimSchMan06JPB}, where the
instantaneous decay rates $\gamma_0(N_b)$ at various populations $N_b$ were used to
predict the time evolution of the quasi-bound population through the numerical
integration of Eq.~(\ref{eq:dNdt}).

\begin{figure}[t]
\begin{center}
\includegraphics[width=0.9\linewidth]{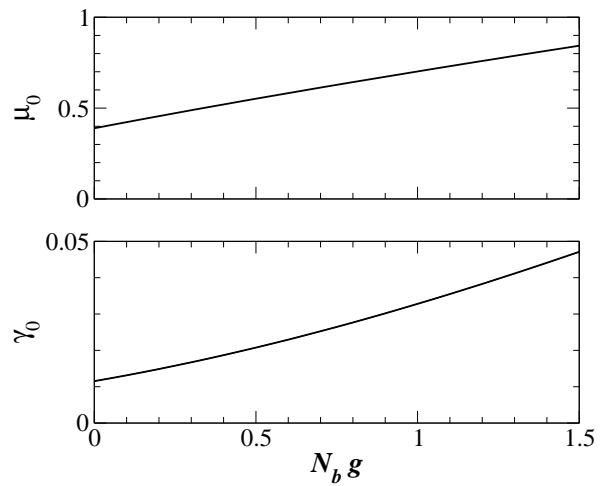}
\caption{
  Chemical potential $\mu_0$ and decay rate $\gamma_0$ of the quasi-bound state
  within the double barrier potential, calculated as
  a function of $N_b g$ with $N_b$ the population of the quasi-bound state
  and $g$ the effective one-dimensional interaction strength.
  In practice, $\mu_0$ and $\gamma_0$ were computed at $30$ equidistant values of
  $N_b g$ within $0 \leq N_b g \leq 1.5$, and cubic interpolation was employed to
  obtain intermediate values of $\mu_0$ and $\gamma_0$ for the self-consistent
  solution of Eq.~(\ref{eq:NBit}).
  $\mu_0$, $\hbar \gamma_0$, and $g / \sigma$ are given in ``natural'' energy units of 
  $\hbar \omega $.
  \label{fg:muga}
}
\end{center}
\end{figure}

In analogy with the noninteracting case, we now replace
$\gamma_\mu \to \gamma_0\left( |B|^2 \right)$ in Eq.~(\ref{eq:trans}), which approximately
interpolates between the decay rate of the weakly populated quasi-bound state
at $\mu = \mu_0(0)$ (which is naturally given by $\gamma_0(0)$) and the decay rate near 
maximum of the shifted resonance peak.
Using this approximation, the equation for the transmission coefficient reads
\begin{equation}
  T(\mu) \simeq \frac{\left[\hbar \gamma_0(N_b) / 2\right]^2}
  {\left[\mu - \mu_0(N_b)\right]^2 + \left[\hbar \gamma_0(N_b) / 2\right]^2},
  \label{eq:transB}
\end{equation}
where the quasi-bound population $N_b$ implicitly depends, via
Eqs.~(\ref{eq:B0}) and (\ref{eq:Source}), on the incident chemical potential
$\mu$ and the incident current $j_i = \hbar k |\alpha|^2 / m$ according to
\begin{equation}
  N_b(t) = \frac{\hbar \gamma_0(N_b) / 2}
  {\left[\mu - \mu_0(N_b)\right]^2 + \left[\hbar \gamma_0(N_b) / 2\right]^2} \hbar j_i \, .
  \label{eq:NBit}
\end{equation}
As in the corresponding decay problem
\cite{CarHolMal05JPB,SchPau06PRA,WimSchMan06JPB}, we now need to know the
instantaneous chemical potentials $\mu_0(N_b)$ and decay rates $\gamma_0(N_b)$ at
given quasi-bound populations $N_b$ in order to calculate solutions of this
set of equations.
We apply for this purpose a real-time propagation method which is based on the
numerical integration of the ``homogeneous'' time-dependent Gross-Pitaevskii
equation (i.e., without the inhomogeneous source term) in presence of
absorbing boundaries.
Starting from an appropriate initial condensate wave function (which should
approximate quite well the resonance state to be calculated), and
renormalizing the wave function after each propagation step to satisfy the
condition
\begin{equation}
  \int_{-a}^a |\psi(x)|^2 dx = N_b \label{eq:renorm}
\end{equation}
within the quantum dot, one indeed obtains, after a sufficiently long
propagation time, convergence towards the lowest decaying state of the system.
The scaling factor that is needed to perform the renormalization
(\ref{eq:renorm}) gives then rise to the decay rate $\gamma_0 = \gamma_0(N_b)$ of
the quasi-bound state, while the chemical potential $\mu_0 = \mu_0(N_b)$ of the
decaying state can be extracted from the expectation value of the nonlinear
Gross-Pitaevskii Hamiltonian.
In practice, it is sufficient to compute $\mu_0$ and $\gamma_0$ in this way for the
equidistant values $N_b g = 0, 0.05, 0.1 \ldots$ of the population $N_b$, and to
use cubic interpolation in order to determine intermediate values of $\mu_0$ and
$\gamma_0$.

\begin{figure}[t]
\begin{center}
\includegraphics[width=\linewidth]{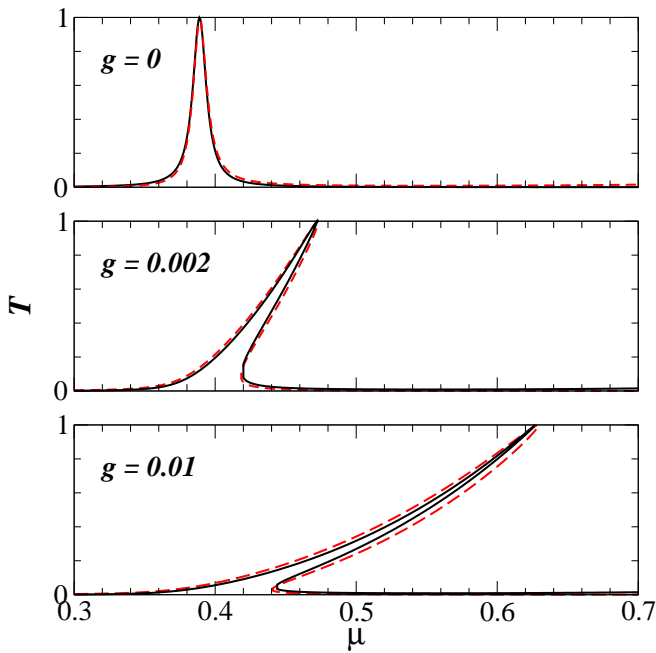}
\caption{(color online)
  Transmission spectra of the double barrier potential at
  $g = 0$ (upper panel), $g = 0.002$ $\hbar \omega a_{\perp}$ (middle panel), and $g =
  0.01$ $\hbar \omega a_{\perp}$ (lower panel).
  The solid line shows the transmissions of all scattering states, calculated
  by the ``stationary'' method based on Eq.~(\ref{160}), that exist at the
  incident current $j_i=1\, \omega$ of the matter-wave beam.
  The dashed line is obtained from self-consistent solutions of
  Eq.~(\ref{eq:NBit}) at $j_i= 1\, \omega$, which are inserted 
  in the expression (\ref{eq:transB}) for the nonlinear transmission
  coefficient.
  The good agreement confirms the one-to-one correspondence between
  quasi-bound states of the atomic quantum dot and resonance peaks in the
  transmission spectrum ($\mu$ in units of $\hbar\omega$)
  \label{fg:trans}
}
\end{center}
\end{figure}

With this information, the possible self-consistent values of the quasi-bound
population can be computed by applying a numerical root-search method to
Eq.~(\ref{eq:NBit}) at given chemical potential $\mu$ and given incident current
$j_i$.
The resulting occupation numbers $N_b$ are then inserted in the expression
(\ref{eq:transB}) for the transmission coefficient.
As shown in Fig.~\ref{fg:trans}, a distorted resonance peak is then obtained
for $g > 0$.
Apart from a slight overestimation of the peak width, this peak agrees quite
well with the peak structure that would be formed through the transmission
coefficients of all possible stationary scattering states at the above
incident density.
This ultimately confirms the one-to-one correspondence between quasi-bound
states of the atomic quantum dot and resonance peaks in the transmission
spectrum.

It is worthwhile to note that self-consistent solutions of the quasi-bound
populations can also be found in a different way, namely by iteratively
inserting approximate expressions for $N_b$ into the right-hand side of
Eq.~(\ref{eq:NBit}) starting with $N_b = 0$.
This approach would effectively mimic the quasi-stationary propagation
of a Bose-Einstein condensate through the initially empty quantum dot.
In agreement with the time-dependent propagation approach based on the
inhomogeneous Gross-Pitaveskii equation (see Sec.~\ref{sec2C}), only the 
\emph{lowest branch} of the distorted resonance peak is populated in this way.
This again underlines that the framework used in this section is intrinsically
suited to take into account time-dependent effects and might therefore be used
to predict the outcome of specific propagation processes.

\section{Conclusion}\label{sec4}
We have presented analytical and numerical results for steady and
time-dependent flows of repulsively interacting Bose condensed atoms through
mesoscopic waveguide structures.
To this end, we described a theoretical framework that is suitable to study
transport and scattering processes in the 1D mean-field regime.
In this context we introduced a non-perturbative method to extend the concept
of transmission and reflection coefficients to nonlinear wave equations.
On the other hand, to predict the behavior of the condensate flow under
realistic experimental conditions, it is necessary to study time-dependent
transport processes.
We developed for this purpose a numerical method based on integrating the
time-dependent \GP in presence of a source term that simulates the coupling of
the waveguide to a reservoir from which  a quasi-stationary flow of condensate
is smoothly released into the guide.

The approach was first applied to the transport through a single quantum point
contact, where we found as a main result that an increasing  nonlinearity
leads to a distinct reduction of the transmission.
Much more complex behavior was found for the condensate flow through a double
barrier potential.
Here, the atom-atom interaction induces a bistability phenomenon of the
transmitted flux in the vicinity of resonances, which manifests as a strong
distortion of the transmission peaks.
By means of the time-dependent integration scheme, we demonstrated that
resonant transport will consequently be suppressed in a realistic propagation
process.
However, as we showed in Ref.~\cite{PauRicSch05PRL}, a suitable variation of
the external potential during the propagation process can enhance the flow to
reach a near-resonant state on finite time scales. 
Finally, an analytical description of the transport problem through the double
barrier was developed, which establishes a clear link between the nonlinear
signatures of the transmission spectra and the properties of the
self-consistent quasi-bound states of the quantum dot.
Similar results were recently obtained in Ref.~\cite{RapKor07} as well.

Our numerical approach based on the inhomogeneous time-dependent
Gross-Pitaevskii equation can be straightforwardly generalized to describe 
scttering processes in multidimensional geometries.
It can certainly be applied also to more complex scattering potentials,
involving more than two barriers.
In that case, however, we do not expect that the calculation always converges
towards a stationary scattering state, even if the source amplitude in the
inhomogeneous Gross-Pitaevskii equation is varied on a very long time scale.
This was demonstrated in our study on the transport of Bose-Einstein
condensates through one-dimensional disorder,
where we found that randomly generated disorder potentials of finite range
will generally give rise to permanently time-dependent scattering processes at
finite interaction, as long as the length of the disorder region exceeds a
critical interaction-dependent value \cite{PauO05PRA,Pau07PRL}.
Interestingly, this cross-over between quasi-stationary and time-dependent
scattering, arising for disorder samples with lengths below and above this
critical value respectively, correlates with a transition from an exponential
(Anderson-like) to an algebraic decrease of the average transmission with the
sample length \cite{PauO05PRA}, which indicates that the depletion of the
condensate during the propagation process might play a prominent role there.
We note in this context that the effect of depletion can to a certain extent
be accounted for within the framework of our approach, namely through the
implementation of the microscopic quantum dynamics approach introduced by
K\"ohler and Burnett \cite{Koe02PRA} in combination with an external source
\cite{Tom}.

The results that were obtained in this work are related to other fields of
nonlinear physics as well, such as nonlinear optics \cite{Vau96PRA} and the
electronic transport through quantum wells \cite{Goldman87,Azbel99}, where
similar observations on resonant transport were made.
In the context of Bose-Einstein condensates, the realization of a
quasi-stationary flux of interacting matter waves though scattering
potentials that are defined on microscopic length scales still represents a
formidable experimental challenge.
There are, however, promising advances in this direction, such as the
atom-laser-like injection of a condensate into an optical waveguide
\cite{Gue06PRL} as well as the scattering of a stationary condensate in
presence of a moving obstacle \cite{Engels}.
Such advances should, in combination with detection techniques for single 
atoms \cite{Teper06,Haase06} (which would allow one to measure very low
transmissions), make it possible to experimentally investigate the role of
interaction in mesoscopic transport processes from a new perspective, namely
the one of cold bosonic atoms.

\section*{Acknowledgments}
It is a pleasure to thank J\'oszef Fort\'agh, Hans-J\"urgen Korsch, Patricio
Leboeuf, Nicolas Pavloff, Kevin Rapedius, Dirk Witthaut, and Carlos Viviescas
for fruitful and inspiring discussions.
Financial support by the Alexander von Humboldt Foundation, by the
Bayerisch-Franz\"osisches Hochschulzentrum, by the Deutsche
Forschungsgemeinschaft, and through the Bayerisches Elitef\"orderungsgesetz is
gratefully acknowledged.

\begin{appendix}

\section{}\label{Appendix1}
In this appendix we describe the numerical integration procedure of the
time-dependent Gross-Pitaevskii equation and the implementation of the source
term into this integration scheme.
We consider the equation of  motion (in the following we set for simplicity
$\hbar=1$, $m=1$)
\begin{equation}\label{A10}
  i\frac{\partial}{\partial t}\psi(x,t) = H(x,t)\psi(x,t),
\end{equation} 
with the effective nonlinear Hamiltonian
\begin{equation}\label{A20}
 H(x,t)\equiv -\frac{1}{2}~\frac{\partial^2}{\partial x^2}+V(x)+g \vert \psi(x,t) \vert^2,
\end{equation} 
which we want to integrate for a given initial state $\psi(x,t_0)$ of the
condensate.
In order to compute the time evolution of the condensate wave function 
$\psi(x,t)$ for $t > t_0$, we subdivide the time interval $t-t_0$ into $n$
discrete time steps of the size $\Delta t=(t-t_0)/n$, and use an implicit
Crank-Nicholson integration scheme \cite{NumRec} to propagate the wavefunction
from one time step to the next one.
The effective time evolution operator $\mathcal{U}$ for one discrete time step
$\Delta t$ is then given by \cite{Ames} 
\begin{equation}\label{A30}
{\mathcal{U}}(t+\Delta t,t)\equiv \frac{1}{1+\frac{i}{2}H(x,t)\Delta t}\left[
  1-\frac{i}{2}H(x,t)\Delta t \right].
\end{equation}
The representation (\ref{A30}) of $\mathcal{U}$ is unitary and thus
conserves the norm of the wave function $\psi$.
The implicit integration scheme for the wave function reads then
\begin{equation}\label{A40}
\left(1+\frac{i\Delta t }{2} H \right)\psi(x,t+\Delta t)=
\left(1-\frac{i \Delta t}{2} H \right)\psi(x,t).
\end{equation}

We expand the wave function on a discrete lattice with $N$ lattice sites by
introducing the grid basis 
\begin{equation}\label{A50}
\chi_j\equiv
\begin{cases}
1 : x_j-\frac{1}{2}\Delta x \leq x < x_j+\frac{1}{2}\Delta x
 \\
0 : \text{otherwise},
\end{cases}
\end{equation}
with $\Delta x \equiv (x_{min} - x_{max})/N$. Here, $x_{min}$ and $x_{max}$ are the
boundaries of the finite grid.
The wave function then reads 
\begin{equation}\label{A60}
\psi(x,t_n)=\sum\limits_{j=1}^N \psi_j^n\chi_j,
\end{equation}
where $\psi^n_j  \equiv \psi(x_j,t_n)$ is value of the wave function at the position
$x_j$ of the $j$'th lattice site (the index $n$ labels the discrete times,
$t_n=t_0 + n \Delta t$).
Using the finite-difference representation for the kinetic part of $H(x,t)$,
we find  
\begin{eqnarray}\label{A70}
\lefteqn{\left(1\pm \frac{i \Delta t}{2} H \right)\psi(x_j,t_n)\simeq\psi_j^n\pm \frac{i\Delta t}{2} \times } 
\nonumber \\
& & \times \left[-\frac{\psi_{j+1}^n-2\psi_{j}^n+\psi_{j-1}^n}{2\,\Delta x^2}+
  V_{j}\psi_j^n+g\vert \psi_j^n\vert^2 \psi_j^n\right] \qquad
\end{eqnarray}
with $V_j \equiv V(x_j)$.
By introducing $\vec{\psi}^n=\left(\psi_1^n...\psi_j^n...\psi_N^n \right)^T$,
the lattice representation of Eq.~(\ref{A40}) finally reads 
\begin{eqnarray}\label{A80}
{\bf D}_2 \vec{\psi}^{n+1}={\bf D}_1\vec{\psi}^n \quad
\Leftrightarrow \quad \vec{\psi}^{n+1}= {\bf D}_2^{-1}{\bf D}_{1}\vec{\psi}^n\, ,
\end{eqnarray}
where we define 
\begin{equation}\label{A90}
{\bf D}_1\equiv \left[\left(1-\frac{i \Delta t}{2} H \right)  \right], \ {\bf D}_2\equiv
\left[\left(1+\frac{i \Delta t}{2} H \right)  \right],
\end{equation}
and the $N\times N$ matrix representation of ${\bf D}_{1,2}$ reads
\begin{equation}\label{A100}
{\bf D}_{1,2}=
\begin{pmatrix}
 \ddots& \ddots& \ddots& & & &\\
 & \pm\alpha & 1\mp\beta_{j-1} & \pm\alpha & & & \\
 & &\pm\alpha &1\mp\beta_j &\pm\alpha  & & \\
 & & &\pm\alpha &1\mp\beta_{j+1} &\pm\alpha &\\
 & & & & \ddots& \ddots &\ddots \nonumber
\end{pmatrix},
\end{equation}
with
\begin{eqnarray}\label{A110}
\alpha\equiv\frac{i \Delta t }{4 \Delta x^2}, \ \
\beta_j\equiv \frac{i \Delta t}{2} \left(\frac{ 1}{\Delta x^2}+ V_j + g\vert \psi_j^n\vert^2\right).
\end{eqnarray}
Hence, the integration of Eq.~(\ref{A10}) reduces to the solution of a
system of linear equations with a tridiagonal matrix.

So far, our integration scheme uses the value of $\psi^n$ at the beginning of the
integration step.
This neglects the fact that the effective Hamiltonian (\ref{A20}) is
implicitly time-dependent due to the presence of the nonlinear term 
$g\vert \psi(x,t) \vert^2$. Thus, it would be appropriate to use a more precise
estimate for this nonlinear term, which is somehow averaged over the timestep
$\Delta t$ leading from $t_n$ to $t_{n+1}$.  
This problem can be handled by using a predictor-corrector-like scheme which
was already successfully applied in \cite{Cerbo}. 
In this scheme, each integration step is done twice: 
First, we propagate the wave function from time $t_n$ to time $t_{n+1}$ 
using $\psi^n$ in the nonlinear term, in order to obtain a {\em predicted} wave
function $\tilde \psi^{n+1}$. Then, we repeat this integration step but using now
the averaged value $\frac{1}{2}[\psi^n + \tilde\psi^{n+1}]$ in the
nonlinear term, yielding a {\em corrected} wave function $\psi^n$.  

Now we consider the presence of the source term.
The equation of motion reads therefore
\begin{equation}\label{A115}
  i\frac{\partial}{\partial t}\psi(x,t) = H(x,t)\psi(x,t)+S(t)\exp(-i\mu t)\, \delta(x).
\end{equation} 
Working with a grid representation of the wave function, it is convenient to
approximate the $\delta$-function by 
\begin{eqnarray}\label{A120}
R(x)=\frac{1}{\Delta x}\left[\Theta(x + \Delta x / 2)  -  \Theta(x - \Delta x /2) \right],
\end{eqnarray}
where $\Theta$ is the Heavyside step function.
Before including the source term to the finite difference scheme, we estimate
the error that is introduced by this approximation. 
To this end, we study the steady-state solutions of the wave equation 
\begin{equation}\label{A130}
 i  \frac{\partial\psi}{\partial t}- \left[ -\frac{1}{2}
 \frac{\partial^2 }{\partial x^2} + V_{\|}(x) +g\vert\psi\vert^2 \right]
\psi =  S_{0}~R(x) \; ,  
\end{equation}
that are obtained in the limit $t\to \infty$.
The Green function that is associated with the stationary equivalent of
Eq.~(\ref{A130}) is given by
\begin{equation} \label{A140}
G(x-x')=\frac{S_{0} }{i k }\;e^{i k \vert x - x'\vert}
\end{equation}
with $k= \sqrt{2 (\mu  -  gn)}$ (see Sec.~\ref{sec2D}, Eq.~(\ref{370})$\,$).
Hence, the ansatz
\begin{equation} \label{A150}
\psi_{_R}(x)=\int_{-\infty}^{+\infty}dx'\, \frac{S_{0} }{i k }e^{i k \vert x-x' \vert} \, R(x')
\end{equation}
yields a solution $\psi_{_R}(x)$ of Eq.~(\ref{A130}).
Evaluating this integral yields
\begin{equation}\label{A160}
  \psi_{_R}(x)=\frac{2 S_{0} }{i k^2 \Delta x}\left\{ \begin{array}{r@{\; : \;}l} 
    \displaystyle e^{-ikx}\sin(k\Delta x/2) & \displaystyle x< - \frac{\Delta x}{2}  \\ 
    \displaystyle 1-e^{ik\Delta x/2}\cos(kx) & \displaystyle |x| <\frac{\Delta x}{2}\\ 
    \displaystyle e^{ikx}\sin(k\Delta x/2) & \displaystyle x>\frac{\Delta x}{2}  
  \end{array} \right.
\end{equation}
which converges towards Eq.~(\ref{A140}) in the limit $\Delta x \to 0$.
The result~(\ref{A160}) can serve as an estimate for the relative error
$\mathcal F$ that is done by approximating $\delta (x)$ with $R(x)$: we obtain
\begin{equation}\label{A170}
{\mathcal F}=1-\frac{2  \sin(k\Delta x/2) }{k \Delta x }\simeq \frac{k^2 \Delta x^2}{24} \
\text{if}  \ \frac{k\Delta x}{2}\ll 1.
\end{equation}
The relative error therefore scales quadratically with the grid spacing $\Delta x$
and becomes negligible for reasonably small values of $\Delta x$.

\begin{figure}[ptb]
\centering
\includegraphics[width=0.9\linewidth,angle=0]{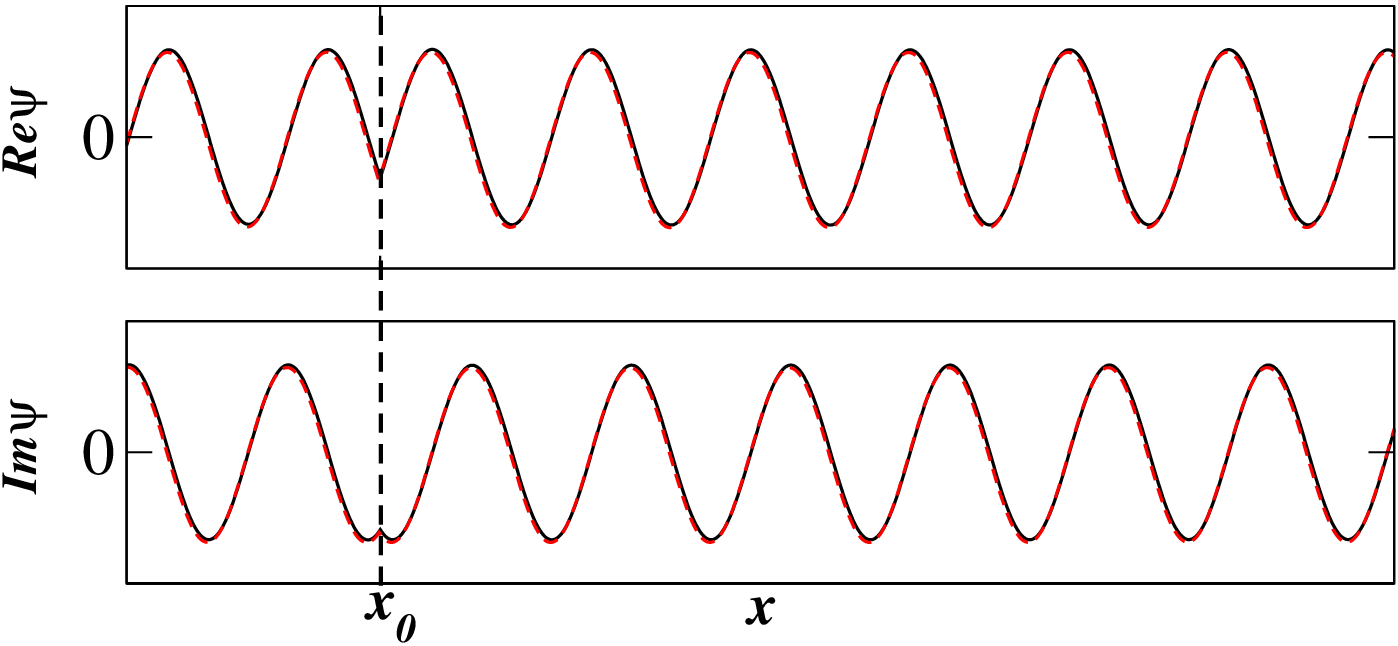}
\\[4mm]
 \caption{\label{Fig_Quelltest} (color online) Real and imaginary parts 
   (black solid lines) of the steady-state plane-wave solution
   obtained by integrating the time-dependent Gross-Pitaevskii equation with
   the numerical source term (\ref{A120}) using the time step 
   $\Delta t= \hbar / (50 \mu)$ and the grid spacing $\Delta x = \lambda / 20$ with $\lambda=2\pi / k$ the
   wavelength of the condensate.
An excellent agreement with the exact analytical result (\ref{A130}) 
(red dashed lines) is found. The source is located at the position $x=x_0$.
}
\end{figure}

The above considerations justify the implementation of the source term at the
position $x_{j'}$ through the discretized form
\begin{equation}\label{A180}
S_j^n=S(t_n)\exp({-i \mu t_n}) \, \delta_{j,j'} \, ,
\end{equation}
where $\delta_{j,j'}=1$ if $j=j'$ and $0$ otherwise.
In the presence of the source term, Eq.~(\ref{A80}) is modified and reads 
\begin{eqnarray}\label{A190}
{\bf D}_2 \vec{\psi}^{n+1} + \vec{b}^n ={\bf D}_1\vec{\psi}^n \Leftrightarrow \vec{\psi}^{n+1}= 
{\bf D}_2^{-1}({\bf D}_1\vec{\psi}^n-\vec{b}^n) \nonumber,
\end{eqnarray}
where the components of the vector $\vec{b}^n$ are given by
\begin{eqnarray}\label{A200}
b_j^n=\frac{i \Delta t}{2}\left(S_{j'}^n + S_{j'}^{n+1} \right)\, \delta_{j,j'}.
\end{eqnarray}
In Fig.~\ref{Fig_Quelltest}, we compare the exact result (\ref{A130}) to
the numerically computed plane-wave solution that is obtained in the limit  
$t\to \infty$ by simulating the gradual filling of a waveguide without scattering
potential, $V_{\|}(x)\equiv 0$.
Indeed, we find an excellent agreement between the numerical result and the
exact plane-wave solution (\ref{A130}) if we choose, e.g., $\Delta x = \lambda / 20$ 
(with the wavelength $\lambda=2\pi / k$) and $\Delta t= \hbar / (50 \mu)$. 

It is worthwhile to mention that in the presence of strong nonlinearities 
(for values of $g$ considerably larger than in this paper) and strong
backreflection, a nonlinear back-action between the reflected matter 
wave and the source term can occur.
As a consequence, the transmitted current depends not only on the source
amplitude $S_0$ but also on the position of the source.
In such a situation, it is advisable to implement the adiabatic transition
scheme that is displayed in Fig.~\ref{Fig_Adpass} where $g$ vanishes in the
far-upstream region. By positioning the source term there and by choosing a
sufficiently large transition region, one can avoid this nonlinear back-action
and ensure that the wave function is adiabatically conveyed from a
linear wave to a nonlinear scattering state obeying the \GP.

\section{}\label{Appendix2}
In the numerical treatment of time-dependent scattering processes in open
quantum systems, one often encounters the problem of defining physically
meaningful boundaries at the edges of the computational domain.
The naive, straightforward expansion of the wave function on a finite spatial
grid generally leads to an artificial backscattering of the wave function
from the boundaries of the grid, which makes it impossible to simulate
infinitely extended scattering states.
This problem can be circumvented by introducing complex absorbing potentials
in the vicinity of the grid boundaries (see, e.g., Ref.~\cite{MoiO04JPB}),
which should be designed such that they absorb the outgoing flux as best as
possible without affecting the dynamics inside the scattering region.
An alternative method, which was introduced by Shibata for the linear
Schr\"odinger equation \cite{Shibata}, consists in the definition of  
\emph{absorbing boundary conditions} (ABC) at the edges of the grid, which are
formulated in order to perfectly match outgoing plane waves with a specified
dispersion relation.
This method is particularly suited for quasi-stationary propagation processes
where the outgoing part of the wave function is well described by a plane
monochromatic wave.
We show in this Appendix how this approach can be numerically implemented, and
how the effect of a moderate nonlinearity in the Gross-Pitaevskii equation can
be taken into account.

We first discuss the absorbing boundary conditions for the Schr\"odinger
equation (with $\hbar=1$ and $m=1$)
\begin{equation}\label{ABC10}
i \frac{\partial}{\partial t}\psi(x,t)=\left(-\frac{1}{2  }\frac{\partial^2}{\partial x^2}+V_e\right)\psi(x,t),
\end{equation}
where $V_e$ is a constant potential which is independent of the position $x$.
This equation admits plane-wave solutions $\psi(x,t)= Ae^{-i(\mu t - k x) }$ satisfying the
dispersion relation 
\begin{equation}\label{ABC20}
 k =  \pm   \sqrt{2  ( \mu - V_e )}.
\end{equation}
The ``$+$'' and ``$-$'' branches of Eq.~(\ref{ABC20}) correspond to plane
waves that propagate to the right- and left-hand side, respectively.
Thus, the ABC should satisfy the dispersion relation given by the ``$+$''
branch of Eq.~(\ref{ABC20}) at the right boundary and the ``$-$'' branch at
the left boundary of the grid. 

\begin{figure}[ptb]
\centering
\includegraphics[width=0.9\linewidth,angle=0]{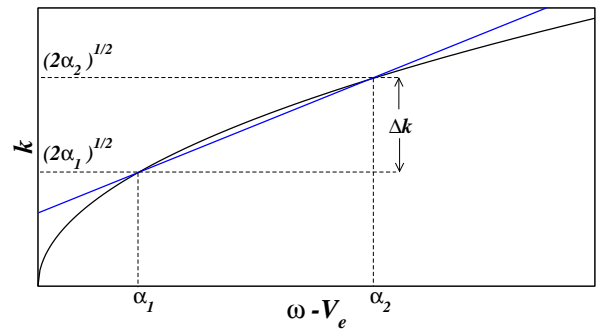}
\\[4mm]
 \caption{\label{Fig_Disprel} (color online) The positive branch of the
   dispersion relation of a plane wave (black line) is approximated by a
   linear function (straight blue line).
   The parameters $\alpha_1, \alpha_2$ are chosen such that the wave numbers of the
   plane waves to be absorbed lie within the momentum interval $\Delta k$.
}
\end{figure}

We derive now so called ``one-way wave equations'' on the basis of the
dispersion relation (\ref{ABC20}), which we will implement at the boundaries
of the grid and which locally allow for wave propagation only in the outgoing
direction.
To this end, we make use of the duality relations
\begin{equation}\label{ABC30}
\frac{\partial}{\partial t} \Longleftrightarrow   - i \mu, \quad \frac{\partial}{\partial x} \Longleftrightarrow i k
\end{equation}
which is going to be inserted into the dispersion relation (\ref{ABC20}).
Unfortunately Eq.~(\ref{ABC20}) is nonlinear in $\mu$ and cannot be
straightforwardly converted into a linear differential equation. 
To circumvent this problem, we approximate Eq.~(\ref{ABC20}) in the vicinity
of the chemical potential of the wave to be absorbed by the linear function 
\begin{equation}\label{ABC40}
 k = \pm \frac{\sqrt{2  \alpha_2} -\sqrt{2  \alpha_1}}{\alpha_2-\alpha_1}~\mu 
~\pm  \frac{\alpha_2\sqrt{2  \alpha_1} -\alpha_1\sqrt{2  \alpha_2}}{\alpha_2-\alpha_1}
\end{equation}
(see Fig.~\ref{Fig_Disprel}).
The parameters $\alpha_1, \alpha_2$ are chosen such that Eq.~(\ref{ABC40}) is a good
approximation to the dispersion relation (\ref{ABC20}) within the interval 
$ \Delta k\equiv\sqrt{2  \alpha_2}-\sqrt{2  \alpha_1}\,$ around the central wave number 
$\frac{1}{2} ( \sqrt{2  \alpha_2}+\sqrt{2  \alpha_1})$.
By use of the duality relations (\ref{ABC30}), Eq.~(\ref{ABC40}) is
transformed into the one-way wave equation
\begin{eqnarray}\label{ABC50} 
i\frac{\partial\psi}{\partial t}&=&\left(-i \frac{1}{g_1}\frac{\partial}{\partial x} + V_e -
  \frac{g_2}{g_1}\right) \psi,
\end{eqnarray}
with
\begin{eqnarray}\label{ABC51} 
g_1&\equiv& \pm  \frac{\sqrt{2 \alpha_2}-\sqrt{2  \alpha_1}}{\alpha_2-\alpha_1}, \nonumber \\
g_2&\equiv& \pm \frac{\alpha_2\sqrt{2  \alpha_1}-\alpha_1\sqrt{2  \alpha_2}}{\alpha_2-\alpha_1}.
\end{eqnarray}
Implementing these one-way wave equations at the boundaries of the grid (see
below) leads to a very good absorption of plane waves with wave numbers $k$
satisfying $\sqrt{2  \alpha_1} \lesssim k \lesssim \sqrt{2  \alpha_2}$.
In Ref.~\cite{Shibata} it was demonstrated that also wave packets of the form 
$\psi\equiv \sum_j A_i\exp(ik_jx)$ can be absorbed if all wave numbers in this
superposition lie within the above interval.

It is straightforward to see that the one-way equations~(\ref{ABC50}) absorb
plane waves also in the presence of the nonlinear term $g|\psi|^2$. 
This is evident for the special case of a constant density:
$\psi(x,t)=\sqrt{n}\exp(-i\mu t/\hbar \pm ikx)$ with the dispersion relation
\begin{eqnarray}\label{ABC60} 
k=\pm \sqrt{2( \mu - g n)}.
\end{eqnarray}
is obviously a solution of the Gross-Pitaevskii equation.
A comparison of Eq.~(\ref{ABC60}) with Eq.~(\ref{ABC20}) reveals that the term
$gn$ can be identified as a constant effective potential. Hence we set 
$V_e\equiv gn$ for a proper absorption of the plane wave. 

We now generalize this result for plane waves whose parameters are slowly
varying in time and position.
This case is of high relevance for our work since the gradual filling of the
guide with matter waves leads to the population of a scattering state whose
outgoing parts, which have to be absorbed at the boundaries of the grid,
exhibit slowly varying amplitudes and phases.
We consider
\begin{eqnarray}\label{ABC70} 
\psi(x,t)=A(x,t)e^{-i\mu t \pm i S(x,t)}
\end{eqnarray}
where $A(x,t)$ and $S(x,t)$ represent the local amplitude and phase,
respectively, of the wave function.
Locally, at position $x=x_0$, we can expand the phase according to
\begin{eqnarray}\label{ABC80} 
S(x,t)=S(x_0,t)+k(x_0,t)(x-x_0)+{\mathcal O}[(x-x_0)^2]
\end{eqnarray}
with $k(x_0,t)\equiv \partial_x S(x,t)|_{x=x_0}$
In the limiting case where $A(x,t)$ and $S(x,t)$ vary on time and length
scales that are considerably larger than $1/\mu$ and $1/k(x_0,t)$, respectively 
(for $x \simeq x_0$ and for all times $t$), Eq.~(\ref{ABC70}) locally takes the
form of a plane wave with a slowly varying amplitude and wave number. 
Under this condition, we find at a given position $x_0$ at any time the local
dispersion relation 
\begin{eqnarray}\label{ABC90} 
k(x_0,t)=\pm \sqrt{2(\mu - g n(x_0,t))}.
\end{eqnarray}
with $n(x_0,t)=|A(x_0,t)|^2$. Hence, $k(x_0,t)$ parametrically depends on $t$
via the condensate density at the position $x_0$ which is supposed to be at
the boundary of the grid. By adjusting the values of $\alpha_1$ and $\alpha_2$ such that
$\sqrt{2  \alpha_1} \lesssim k(x_0,t) \lesssim \sqrt{2  \alpha_2}$ is satisfied for all times $t$, the
wave $\psi$ is absorbed at the edge of the lattice.

\begin{figure}[ptb]
\centering
\includegraphics[width=0.69\linewidth,angle=0]{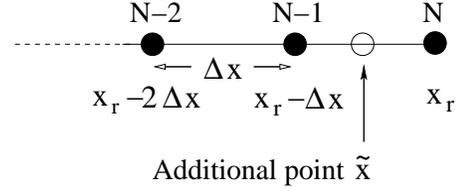}
\\[4mm]
 \caption{\label{Fig_Gitterrand} Sketch of the right lattice boundary. 
   The additional point at position $\tilde x$ allows for a proper
   implementation of the absorbing boundary conditions in a grid
   representation of the wave function.
}
\end{figure}

We now outline how to incorporate the ABC into the lattice representation
(\ref{A60}) of the wave function.
Here we consider exemplarily the right-hand side boundary $x_r=x_N$ of the grid,
where the wave function has to obey Eq.~(\ref{ABC50}) with the upper ($+$)
sign in the definition (\ref{ABC51}) of the prefactors.
The idea is now to replace the equation for the boundary component $\psi_{N}^n$ of
the state vector $\vec{\psi}^n$, i.e.\ the last component in the equation
(\ref{A80}), by the finite-difference version of the one-way wave equation
(\ref{ABC50}).
To this end, we need a finite-difference expression for the derivative 
$\partial_x \psi(x,t)|_{x=x_r}$ at the grid boundary.
Since $x_r$ is the last grid point, this expression can only be obtained in an
asymmetric way with respect to $x_r$, namely through the difference between
$\psi(x_r,t)$ and $\psi(x_r- \Delta x,t)$. This would lead to the equation
\begin{eqnarray}
\lefteqn{\frac{i}{\Delta t}[\psi(x_r, t+\Delta t)-\psi(x_r,
  t)]=\left(V_e-\frac{g_2}{g_1}\right)\psi(x_r, t)} 
\nonumber \\
&&\qquad \qquad \quad - 
\frac{i}{g_1}\frac{\psi(x_r,t)-\psi(x_r- \Delta x,t)}{\Delta x} \qquad \label{ABC95} 
\end{eqnarray}
which was also used in Ref.~\cite{Shibata}.

The asymmetric structure of Eq.~(\ref{ABC95}) introduces a small but
systematic error in the propagation of the wave function, since the value
and the derivate of $\psi$ are, strictly speaking, computed at different
positions, namely at $x_r$ and at the intermediate point 
$\tilde x = x_r - \Delta x / 2$, respectively.
This problem can be circumvented by replacing Eq.~(\ref{ABC95}) with the
analogous equation for the wave function $\psi(\tilde x,t)$ evaluated at this
intermediate point $\tilde x$ (see Fig.~\ref{Fig_Gitterrand}).
There we have
\begin{eqnarray}\label{ABC100} 
\left. \frac{\partial}{\partial x}\psi(x, t)\right|_{x=\tilde x}  \simeq\frac{\psi(x_r,t)-\psi(x_r- \Delta
  x,t)}{\Delta x}\, .
\end{eqnarray}
as ``exact'' (i.e., symmetric) finite-difference expression for the
derivative, and the value of the wave function at this additional point is
obtained through
\begin{eqnarray}\label{ABC110} 
\psi(\tilde x,t) \simeq \frac{1}{2}\left[\psi(x_r,t)+\psi(x_r-\Delta x,t)\right].
\end{eqnarray}
Inserting these expressions (\ref{ABC100}) and (\ref{ABC110}) into 
Eq.~(\ref{ABC50}) leads to a \emph{symmetric} finite-difference equation for
$\psi(x_r,t)$ and $\psi(x_r- \Delta x,t)$ where the value of the wave function at the
auxiliary point $\tilde x$ does not explicitly appear any longer.
In the grid representation, this finite-difference equation reads 
\begin{eqnarray}\label{ABC120} 
  \lefteqn{\frac{i}{2 \Delta t}(\psi_N^{n+1}+\psi_{N-1}^{n+1} -\psi_N^{n} - \psi_{N-1}^{n})= 
  \frac{-i}{g_1 \Delta x}(\psi_N^n-\psi_{N-1}^n)} \nonumber  \\
  && \qquad \qquad \qquad  
  +\frac{1}{2}\left(V_e-\frac{g_2}{g_1}\right) \left(\psi_N^n+\psi_{N-1}^n \right). 
  \qquad \qquad
\end{eqnarray}
Eq.~(\ref{ABC120}) allows for a straightforward incorporation into the matrix
representation (\ref{A90}): the modified matrices ${\bf D}_{1,2}$ read at the
right-hand side edge of the numerical grid 
\begin{eqnarray}\label{ABC130} 
{\bf D}_1&\equiv& 
\begin{pmatrix}
 &\ddots&\ddots &\ddots & & \\
 & & \alpha & 1-\beta_{N-2} & \alpha & \\
 & & &\alpha &1-\beta_{N-1}  & \alpha \\
 & & & &\gamma_3 &\gamma_4
\end{pmatrix},
\nonumber\\[4mm]
{\bf D}_2&\equiv& 
\begin{pmatrix}
 & \ddots & \ddots &\ddots & & \\
 & & -\alpha & 1+\beta_{N-2} & -\alpha &  \\
 & & &-\alpha &1+\beta_{N-1}  & -\alpha \\
 & & & &\gamma_1 &\gamma_2  
\end{pmatrix}
\\&&\nonumber
\end{eqnarray}
where we define
 \begin{eqnarray}
\gamma_1&\equiv&\gamma_2\equiv \frac{i}{2 \Delta t},\nonumber \\
\gamma_3&\equiv& \frac{i}{2 \Delta t}+\frac{i}{g_1 \Delta x}
+\left(V_e-\frac{g_2}{g_1}\right),\nonumber \\
\gamma_4&\equiv& -\frac{i}{2 \Delta t}-\frac{i}{g_1 \Delta x}+\left(V_e -\frac{g_2}{g_1}\right).
\end{eqnarray}

The main cause for artificial backreflection in presence of the 
above boundary conditions comes from the approximate nature of the
finite-difference evaluations (\ref{ABC100}) and (\ref{ABC110}).
Clearly, these approximations become better with decreasing 
grid spacing $\Delta x$, which means that a reduction of the grid spacing 
should lead to a more efficient absorption of the outgoing flux.
In practice, we find for grid spacings of the order of $\Delta x = \lambda / 30$ 
(with $\lambda = 2 \pi / k$ the wavelength of the condensate) that the relative
amplitude of artificial backreflections from the grid boundaries is below
$1\%$.
We note that the amount of backreflection that is accumulated during the
numerical propagation process would, at the same value of the grid spacing 
$\Delta x$, be considerably larger if the asymmetric version (\ref{ABC95})
of the one-way wave equation was used instead of Eq.~(\ref{ABC120}).

\end{appendix}


\begin{thebibliography}{99}
\bibitem{Bemerkung1}
Obviously, numerical convergence towards dynamically unstable scattering
states is not possible.
\bibitem{Bem1}
A multilayer chip geometry might be employed to avoid decoherence and
fragmentation effects that would result from the near vicinity of the
condensate to the wires creating the magnetic waveguide.
\end{thebibliography}
\end{document}